\documentclass[12pt]{article}
\usepackage{graphicx,subfig}
\usepackage{amsmath,amssymb}
\begin{document}

\def\a{\alpha}
\def\b{\beta}
\def\c{\varepsilon}
\def\d{\delta}
\def\e{\epsilon}
\def\f{\phi}
\def\g{\gamma}
\def\h{\theta}
\def\k{\kappa}
\def\l{\lambda}
\def\m{\mu}
\def\n{\nu}
\def\p{\psi}
\def\q{\partial}
\def\r{\rho}
\def\s{\sigma}
\def\t{\tau}
\def\u{\upsilon}
\def\v{\varphi}
\def\w{\omega}
\def\x{\xi}
\def\y{\eta}
\def\z{\zeta}
\def\D{\Delta}
\def\G{\Gamma}
\def\H{\Theta}
\def\L{\Lambda}
\def\F{\Phi}
\def\P{\Psi}
\def\S{\Sigma}

\def\o{\over}
\def\beq{\begin{eqnarray}}
\def\eeq{\end{eqnarray}}
\newcommand{\gsim}{ \mathop{}_{\textstyle \sim}^{\textstyle >} }
\newcommand{\lsim}{ \mathop{}_{\textstyle \sim}^{\textstyle <} }
\newcommand{\vev}[1]{ \left\langle {#1} \right\rangle }
\newcommand{\bra}[1]{ \langle {#1} | }
\newcommand{\ket}[1]{ | {#1} \rangle }
\newcommand{\EV}{ {\rm eV} }
\newcommand{\KEV}{ {\rm keV} }
\newcommand{\MEV}{ {\rm MeV} }
\newcommand{\GEV}{ {\rm GeV} }
\newcommand{\TEV}{ {\rm TeV} }
\def\diag{\mathop{\rm diag}\nolimits}
\def\Spin{\mathop{\rm Spin}}
\def\SO{\mathop{\rm SO}}
\def\O{\mathop{\rm O}}
\def\SU{\mathop{\rm SU}}
\def\U{\mathop{\rm U}}
\def\Sp{\mathop{\rm Sp}}
\def\SL{\mathop{\rm SL}}
\def\tr{\mathop{\rm tr}}

\def\IJMP{Int.~J.~Mod.~Phys. }
\def\MPL{Mod.~Phys.~Lett. }
\def\NP{Nucl.~Phys. }
\def\PL{Phys.~Lett. }
\def\PR{Phys.~Rev. }
\def\PRL{Phys.~Rev.~Lett. }
\def\PTP{Prog.~Theor.~Phys. }
\def\ZP{Z.~Phys. }


\baselineskip 0.7cm

\begin{titlepage}

\begin{flushright}
IPMU12-0008 \\
ICRR-REPORT-604-2011-21
\end{flushright}

\vskip 1.35cm
\begin{center}
{\large \bf A 125\,GeV Higgs Boson and Muon $g-2$\\ in More Generic Gauge Mediation
}
\vskip 1.2cm
Jason L. Evans$^1$, Masahiro Ibe$^{1,2}$, Satoshi Shirai$^{3,4}$ and Tsutomu T. Yanagida$^1$
\vskip 0.4cm
$^1${\it IPMU, TODIAS, University of Tokyo, Kashiwa 277-8583, Japan}\\
$^2${\it ICRR, University of Tokyo, Kashiwa 277-8582, Japan}\\
$^3${\it 
Department of Physics, University of California, Berkeley, CA 94720}\\
$^4$
{\it
Theoretical Physics Group, Lawrence Berkeley National Laboratory, \\Berkeley, CA 94720
}

\vskip 1.5cm

\abstract{
Recently, the ATLAS and CMS collaborations reported exciting hints of a
Standard Model-like Higgs boson with a mass around $125$\,GeV.
A Higgs boson this heavy is difficult to
realize in conventional models of gauge mediation.
Here we revisit the lightest Higgs boson mass in ``more generic gauge mediation,"
where the Higgs doublets mix with the messenger doublets.
We show that a Higgs boson mass around $125$\,GeV can be realized in
more generic gauge mediation models, even for a relatively light gluino mass, $m_{\rm gluino}\sim 1$\,TeV.
We also show that the muon anomalous magnetic moment can be
within $1\sigma$ of the experimental value for these models, even when the Higgs boson is relatively heavy.
We also discuss the LHC constraints and the prospects of discovery.
}
\end{center}
\end{titlepage}

\setcounter{page}{2}

\section{Introduction}
The ATLAS and CMS collaborations collected almost $5\,$fb$^{-1}$ data in 2011\,\cite{ATLAS, CMS}.
A tantalizing hints for a Standard Model-like Higgs boson has
emerged from this data with a mass around $125$\,GeV.
Although these results are not conclusive enough to claim discovery,
such a relatively heavy Higgs boson would have significant impact
on the supersymmetric (SUSY) Standard Model (SSM), if it is indeed
confirmed by further data collection.

In particular, a lightest Higgs boson mass around $125$\,GeV
is very problematic in the minimal SSM (MSSM).
To realize such a heavy Higgs boson in the MSSM,
we need either very large squark masses, $O(10-100)$\,TeV\,\cite{Okada,TU-363},
or a large stop $A$-term with stop squark masses
around a TeV (see recent discussions in Refs.\,\cite{Heinemeyer:2011aa,Arbey:2011ab,Draper:2011aa}).
In the former case, the search for the superparticles at the early LHC is quite
difficult, even if the gauginos are within reach of the LHC
as in Split-Supersymmetry\,\cite{ArkaniHamed:2004fb,Giudice:2004tc,ArkaniHamed:2004yi}
or pure gravity mediation models\,\cite{Ibe:2011aa}
(see also Refs.\,\cite{Ibe:2006de,Moroi:2011ab}).
In this sense, the latter case with large $A$-terms is more interesting for the LHC experiments
where we have a better chance of discovering the superparticles in near future.

It is, however, not easy to find models with large $A$-terms at the low-scale.
The major obstacle in creating a viable model is the
suppression of the $A$-term during renormalization group
running to the low-scale. In gravity mediation models,
for example, the high-scale cutoff is rather larger, i.e. GUT scale or Plank scale. The $A$-terms are renormalization group evolved over many orders of magnitude. 
This prolonged running drastically suppresses the $A$-terms.
To offset this suppression, the $A$-terms must be very large at the high-scale. Other models are worse. For example, minimal gauge mediation has nearly vanishing $A$-term at the messenger scale.

In a recent paper\,\cite{Evans:2011bea}, three of us (J.L.E., M.I. and T.T.Y.)
constructed a class of models termed  ``more generic gauge mediation." In these models, the messenger doublets are mixed with the Higgs doublets without generating flavor changing neutral currents (FCNC) or rapid proton decay.
It was also shown that the desired large $A$-terms can be generated.
The suppression of the $A$-terms due to the renormalization group evolution in these models, however,
is minimal since the messenger scale can be as low as $O(100)$ TeV.
As a result, a relatively heavy Higgs boson was obtained\,\cite{Evans:2011bea}.

In light of the recent results of the ATLAS and CMS collaborations, we revisit the lightest Higgs boson mass in these more generic gauge mediation models.
We also show that the muon anomalous magnetic moment
can be consistent with the experimental value
at the $1\sigma$ level in large regions of parameter space.%
\footnote{See discussions
on the simultaneous explanation of a lightest Higgs boson
mass around $125$\,GeV and the deviation of the muon
$g-2$ in the focus point supersymmetry\,\cite{Feng:2011aa}
in models with extra matter\,\cite{Moroi:2011aa,Endo:2011xq}
and in models with extended gauge interactions\,\cite{Endo:2011gy}.
}
We stress here that the field content of more generic gauge mediation
is the same as in minimal gauge mediation.
It is quite surprising that merely introducing mixing between the Higgs and the messenger doublets
can resolve the tension in gauge mediation models, i.e. a relatively heavy Higgs boson and a large enough muon $g-2$.

The organization of the paper is as follows.
In section\,\ref{sec:MGG}, we review more generic gauge mediation models.
In section\,\ref{sec:Higgs}, we show that a relatively heavy Higgs boson
and a consistent muon $g-2$ can be simultaneously obtained
in more generic gauge mediation models.
In section\,\ref{sec:LHC}, we discuss constraints and prospects of detection for the present model at the LHC experiments.
The final section is devoted to our conclusions and discussions.

\section{More Generic Gauge Mediation}\label{sec:MGG}
\subsection{Higgs-messenger mixing}
Let us briefly review our more generic gauge mediation model which was constructed in Ref.\,\cite{Evans:2011bea}.
In these more generic gauge mediation models,
we allowed the Higgs doublets to mix with the doublet portion of the messenger multiplets
via the superpotential couplings,
\begin{eqnarray}
 W_{\rm mixing} = gZ\bar{\Phi}{\Phi}
 + g'Z\bar{\Phi}_{\bar L} {H}_u
  + g''Z H_d{\Phi}_{\bar L}\ ,
\end{eqnarray}
while the unwanted flavor mixing and proton decay operators are suppressed
(see Ref.\,\cite{Evans:2011bea} for detailed discussion).
Here, we have assume the messengers $(\Phi, \bar{\Phi})$ are a
fundamentals and anti-fundamentals of the minimal grand unified gauge group, $SU(5)$,
and we split the messengers into $\Phi = (\Phi_{D}, \Phi_{\bar L})$
and $\bar\Phi = (\bar{\Phi}_{D}, \bar{\Phi}_{\bar L})$ in accordance with
the MSSM gauge charges.
We also treat the supersymmetry breaking field $Z$ as a spurion which breaks supersymmetry
having the vacuum expectation value,
\begin{eqnarray}
\label{eq:Z1}
 g \vev{Z} = M+ F\theta^2 \ .
\end{eqnarray}
In Ref.\,\cite{Evans:2011bea}, we found four possible classes of gauge mediation which are
consistent with flavor constraints as well as rapid proton decay constraints;
\begin{itemize}
\item No mixings between the messengers and the Higgs pair (i.e. $g' = g'' = 0$).
\item The messenger $\Phi_{\bar L}$ mixes with $H_u$ (i.e. $g' \neq 0$, $g'' = 0$).
\item The messenger $\bar{\Phi}_{\bar L}$ mixes with $H_d$ (i.e. $g' = 0$, $g'' \neq 0$).
\item The messengers $\Phi_{\bar L}$ and $\bar{\Phi}_{\bar L}$ mix with $H_u$ and $H_d$, respectively
(i.e. $g' \neq 0$, $g'' \neq 0$).
\end{itemize}
Each class of models can be realized
with the help of a ``charged"  coupling constant, i.e. the SUSY-zero
mechanism\,\cite{Evans:2011bea} (see also the appendix\,\ref{sec:symmetry}).%
\footnote{
To realize the fourth class of model by the SUSY zero mechanism,
we need at least two pairs of messengers.
A similar model to the fourth class has been considered
based on the framework of extra dimensions\,\cite{Chacko:2001km}. However, these models have more difficulty realizing a $125$ GeV lightest Higgs boson.
}
The first class of models corresponds to conventional gauge mediation.
As emphasized in Ref.\,\cite{Evans:2011bea},
the  second class of models, which was named Type-II gauge mediation,
leads to a peculiar mass spectrum when compared with conventional gauge mediation. Particularly, the lightest Higgs boson mass can be rather large and a mass of $125$\,GeV can be
easily realized even if the gluino mass is relatively light, $m_{\rm gluino}\lesssim 2$\,TeV.
In the following discussion, we concentrate on these Type-II models since we are most
interested in the mass of the lightest Higgs boson.
However, these other two new classes of models will have their own unique spectrum.

Before closing this section, it should be noted that more generic gauge mediation requires messengers that couple to a suprion which has both a scalar expectation value as well as an $F$-term expectation value as in Eq.\,(\ref{eq:Z1}).
To realize such a messenger sector with a stable vacuum,
the origin of the spurion field should be a secondary supersymmetry breaking
field as realized in ``cascade supersymmetry breaking" models\,\cite{Ibe:2010jb,Evans:2011pz}
(see earlier implementations of the cascade supersymmetry breaking
 \cite{Dine:1993yw,Dine:1994vc,Dine:1995ag,Nomura:1997ur,Fujii:2003iw}
 which revived the original ideas of the gauge 
 mediation \cite{Dine:1981za, Dine:1981gu, Dimopoulos:1982gm}
).
In cascade supersymmetry breaking, the size of the primary supersymmetry
breaking is generally much larger than the secondary breaking appearing in Eq.\,(\ref{eq:Z1}).
As a result, the gravitino mass is expected to be not too light, i.e. $m_{3/2}>{\cal O}(100)$\,keV.%
\footnote{In some models of cascade supersymmetry breaking,
a light gravitino can be realized in a non-perturbative limit\,\cite{Ibe:2010jb,Evans:2011pz}.}

\subsection{Soft parameters in Type-II gauge mediation model}
Let us discuss the soft parameters peculiar to Type-II gauge mediation models
where only $H_u$ mixes with the messengers.
The superpotential of Type-II gauge mediation at the messenger scale is given by
\begin{eqnarray}
\label{eq:superpotential1}
 W = gZ\bar{\Phi}\tilde{\Phi}
 + g'Z\bar{\Phi}_{\bar L}  \tilde{H}_u
  + \tilde \mu \tilde{H}_uH_d  + \tilde{y}_{U ij} \tilde{H}_u Q_{Li} \bar{U}_{Rj}\ ,
\end{eqnarray}
where $\tilde\mu$ is a dimensionful parameter,  $\tilde y_{Uij}$
is the usual $3\times 3$ Yukawa coupling matrix.
We have also placed tildes on $H_u$ and $\Phi_{\bar L}$ for later purposes
and have neglected the rest of the MSSM superpotential which is not relevant for our discussion.
The unwanted terms such as $\Phi_{\bar L} Q_L \bar{U}_R$ and $\Phi_{D} Q_L Q_L$
can be forbidden because of the SUSY zero mechanism\,\cite{Evans:2011bea}.%
\footnote{In Eq.\,(\ref{eq:superpotential1}),
we are assuming that the possible Higgs-Messenger mixing in K\"ahler potential
has been eliminated by appropriate field redefinitions.
Such field redefinitions leads to additional terms such as $\tilde{\Phi}_{\bar L} Q_L \bar{U}_R$.
However, they do not cause the unwanted flavor-changing effects since
their flavor structure is aligned with the Yukawa interaction $H_u Q_L\bar{U}_R$.
In the following, we neglect such effects by assuming $g, g'\ll 1$, although
our discussion is not changed too much even for $g,g' = O(1)$.
}
The explicit charge assignments for the SUSY zero mechanism are given in 
the appendix\,\ref{sec:symmetry}.%
\footnote{
The charge assignments defined in Ref.\,\cite{Evans:2011bea}
are incomplete to suppress unwanted terms in the superpotential, while
the ones in the appendix\,\ref{sec:symmetry} completely suppress all the unwanted terms.
}

To elicit the important low-scale phenomenon, we change our field basis by the rotation
\begin{eqnarray}
\left(
\begin{array}{cc}
\tilde \Phi_{\bar L}\\
\tilde H_u
\end{array}
\right)=
\frac{1}{\sqrt{g^2+g'^2}}\left(
\begin{array}{cc}
g  & -g'     \\
g'  & g
\end{array}
\right)
\left(
\begin{array}{cc}
\Phi_{\bar L}\\
H_u
\end{array}
\right)\ .
\end{eqnarray}
In this new basis, the superpotential becomes
\begin{eqnarray}
\label{eq:superpotential2}
 W = \bar{g} Z\bar\Phi {\Phi}  + \mu H_uH_d   + \mu' \Phi_{\bar L} {H}_d + y_{Uij} H_u Q_{Li} \bar{U}_{Rj}
 + y_{Uij}' \Phi_{\bar L} Q_{Li} \bar{U}_{Rj}\ ,
\end{eqnarray}
where the parameters are defined as
\begin{eqnarray}
 \bar{g} &=& \sqrt{g^2 + g'^2}\ ,\quad
 \mu = \frac{g}{\sqrt{g^2 + g'^2}}\tilde\mu\ , \quad
  \mu' = \frac{g'}{\sqrt{g^2 + g'^2}}\tilde\mu\ , \quad\cr
&&  y_{Uij} = \frac{g}{\sqrt{g^2 + g'^2}}\tilde y_{Uij}\ , \quad
  y_{Uij}' = \frac{g'}{\sqrt{g^2 + g'^2}}\tilde y_{Uij} \ .
\end{eqnarray}
This new basis is much better for calculating low scale physics because the only heavy states are clearly $\Phi,\bar\Phi$.
Hereafter, we also change the definition of the spurion in
Eq.\,(\ref{eq:Z1}) by replacing $g$ with $\bar{g}$.
In this basis, the mixing angle between
the Higgs and the messengers doublets is suppressed by $O(\m/M)$, as compared to $O(g'/g)$ in the original.
Since we will consider $g'/g\sim 1$, this basis is better suited for physics below the messenger scale.

It should be noted that the new flavor dependent interactions,
\begin{eqnarray}
 W = y_{Uij}' \Phi_{\bar L} Q_{Li} \bar{U}_{Rj}\ ,
\end{eqnarray}
are not dangerous.
These new flavor dependent interactions are aligned with the MSSM
Yukawa coupling, $y_U$,
and hence, $y_U$ and $y_U'$ can be simultaneously diagonalized.%
\footnote{In this sense, Type-II gauge mediation
is a natural realization of the so called ``minimal flavor violation" scenario (see for example Ref.\,\cite{D'Ambrosio:2002ex}).
}
In the following discussion, we choose the basis where ${\tilde y}_U$ is diagonal and neglect everything except the top Yukawa coupling,
\begin{eqnarray}
\label{eq:top}
W = y_t H_u Q_{L3} \bar{T}_R +  y_t' \Phi_{\bar L} Q_{L3} \bar{T}_R\ .
\end{eqnarray}
Not only are these interactions not dangerous, but it is these new interactions that give Type-II gauge mediation its unique spectrum.

As discussed in Ref.\,\cite{Evans:2011bea}, the newly added interaction in Eq.\,(\ref{eq:top}) leads
to an $A$-terms at the one-loop level,
\begin{eqnarray}
 A_t  = -\frac{3}{32\pi^2} y_t'^2 \frac{F}{M}\frac{1}{x}\log\left(\frac{1+x}{1-x}\right)\ ,
\end{eqnarray}
for the stop and
\begin{eqnarray}
 A_b  = -\frac{1}{32\pi^2} y_t'^2 \frac{F}{M}\frac{1}{x}\log\left(\frac{1+x}{1-x}\right)\ ,
\end{eqnarray}
for the sbottom.
Here, we have defined $x = F/M^2$, and the above results reduce to
\begin{eqnarray}
\label{eq:Aterm2}
 A_t \simeq - \frac{3 y_t'^2} {16\pi^2} \frac{F}{M}\ ,
\end{eqnarray}
for $x \ll 1$.
We see that the one-loop contribution to the $A$-terms can be comparable
to the gauge mediated soft masses squared
\begin{eqnarray}\label{eq:GMSB}
 m_{Q,T}^{2} \simeq \frac{8}{3}\left(\frac{\alpha_3}{4\pi}\right)^2 \frac{F^2}{M^2}\ , \quad (x\ll1 ).
\end{eqnarray}
for $y_t' \simeq 1$.

The soft masses squared of $Q_{L3}$ and $\bar{T}_R$ are also generated
at the one-loop level, and are given
by
\begin{eqnarray}
\label{eq:oneloop1}
\delta m_{Q_3}^2
&=&  \frac{y_t'^2}{32\pi^2}\frac{F^2}{M^2}
\left(
\frac{
(2+x)\log(1+x)
+
(2-x)\log(1-x)
}{x^2}
\right)\ ,
\end{eqnarray}
and
\begin{eqnarray}
\label{eq:oneloop2}
\delta m_{\bar{T}}^2 =  2 \times \delta m_{Q_3}^2 \ .
\end{eqnarray}
It should be noted that these one-loop contributions to the stop squared masses are
negative\,\cite{Evans:2011bea}.
The negative contributions, however, are subdominant
for $x \ll 1$, since they are suppressed by $x^2$
compared to the positive two-loop contributions of gauge mediation.

Besides these one-loop contributions, the newly added interaction leads to
a sizable two-loop contribution to $m_Q^2$,  $m_T^2$ and $m_{H_u}^2$.
Unlike the one-loop contributions to $m_Q^2$ and $m_T^2$,
the two-loop contributions are not suppressed in the limit of $x\ll 1$.
The leading two-loop contributions can easily be extracted from the wave function renormalization by analytic continuation into superspace\,\cite{Giudice:1997ni} leading to
\begin{eqnarray}
\label{eq:Twoloop}
 \delta m_{Q_3}^2 &=& \frac{y_t'^2}{128\pi^4}\left(3y_t'^2 +3y_t^2
 - \frac{8}{3}g_3^2
 -\frac{3}{2} g_2^2
 -\frac{13}{30} g_1^2\right)\frac{F^2}{M^2}\ , \cr
 \delta m_{\bar T}^2 &=& \frac{y_t'^2}{128\pi^4}\left(6y_t'^2  + 6y_t^2
 + y_b^2
 - \frac{16}{3}g_3^2
 -3 g_2^2
 -\frac{13}{15} g_1^2\right)\frac{F^2}{M^2}\ ,\cr
  \delta m_{\bar B}^2 &=& - \frac{y_b^2y_t'^2 }{128\pi^4}\frac{F^2}{M^2}\ , \nonumber \\
  \delta m_{H_u}^2 &=& - 9\frac{y_t^2y_t'^2 }{256\pi^4}\frac{F^2}{M^2}\ , \nonumber \\
    \delta m_{H_d}^2 &=& - 3\frac{y_b^2y_t'^2 }{256\pi^4}\frac{F^2}{M^2}\ ,
\end{eqnarray}
where $y_b$ is the bottom Yukawa coupling constant and $m_{\bar B}^2$
is the soft squared mass of the right-handed sbottom.
The derivation of these results are given in the appendix\,\ref{sec:appA}.

The soft SUSY breaking squared mass for $H_d$
also has a ``tree-level" contribution due to the third term in the superpotential of Eq.\,(\ref{eq:superpotential2}).
By integrating out the messengers, the down-type Higgs $H_d$ gets a tree-level soft squared mass,
\begin{eqnarray}
\label{eq:tree}
   m_{\bar H}^2 = - \mu'^2 \frac{F^2}{M^4-F^2}\ .
\end{eqnarray}
Here, $\mu'$ is assumed to be of the same order of magnitude as the $\mu$-term,
for $g/g'=O(1)$.
This contribution can be important in low scale gauge mediation where $F/ M^2\simeq 1$.
However, as we push up the messenger scale this contribution falls off quickly.
This tree-level mediation does not play an important role
in most of the parameter space we are interested in.

Finally, let us summarize the parameters of type-II gauge mediation models.
It should be noted that the only new interaction is the one given in Eq.\,(\ref{eq:top}),
and hence, $y_t'$ is the only additional parameter not present in conventional gauge mediation models.
That is, type-II gauge mediation model can be parameterized by,
\begin{eqnarray}
 N_5, \quad \Lambda = \frac{F}{M}, \quad M, \quad \tan\beta, \quad y_t'\ , \quad sgn(\mu)\ ,
\end{eqnarray}
where $N_5$ is the effective number of the messenger multiplets, and
$\tan\b$  is the ratio of the two vacuum expectation values of the Higgs doublets.
As discussed in Ref.\,\cite{Evans:2011bea}, the soft terms generated by this single interaction
significantly change the prediction on the lightest Higgs boson mass.

\section{Higgs Boson Mass and Muon $g-2$}\label{sec:Higgs}
\begin{figure}[t]
\begin{center}
\begin{minipage}{.49\linewidth}
  \includegraphics[width=.9\linewidth]{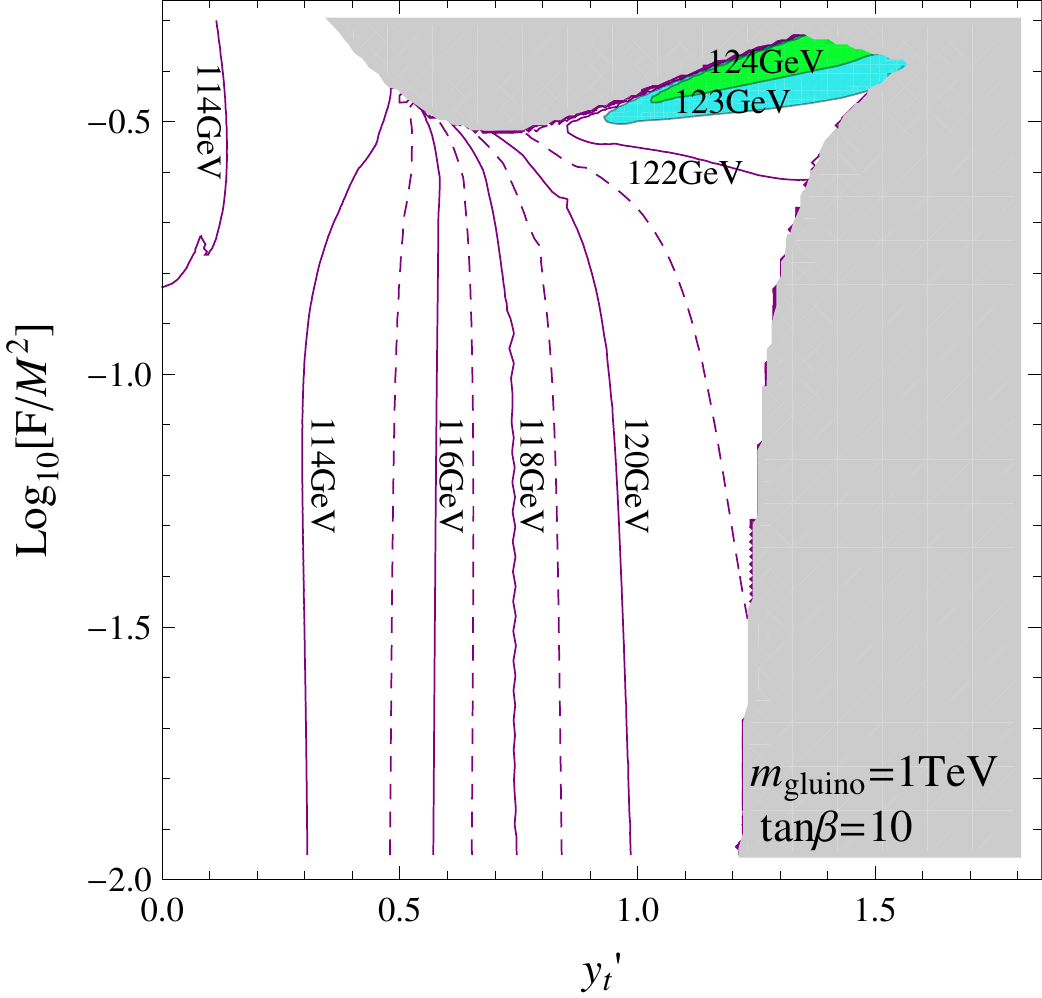}
  \end{minipage}
  \begin{minipage}{.49\linewidth}
  \includegraphics[width=.9\linewidth]{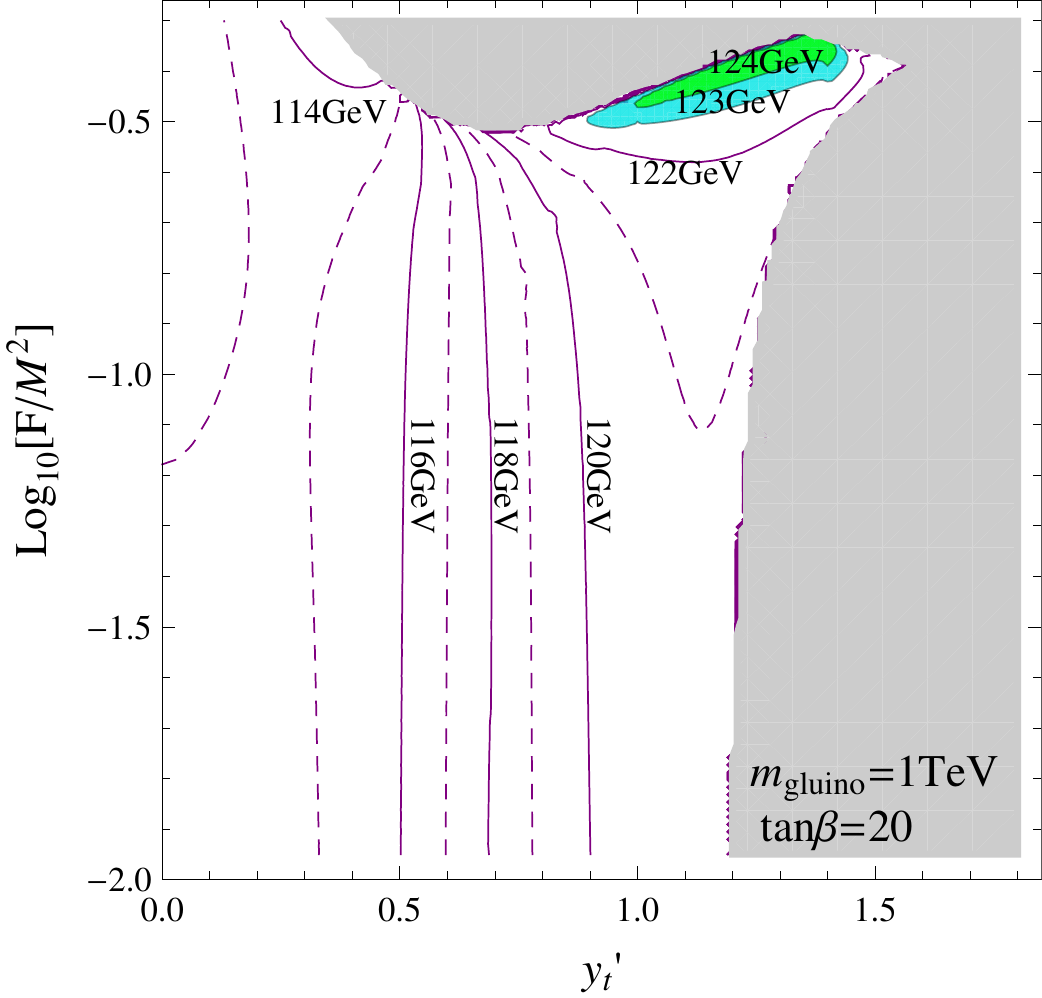}
  \end{minipage}
\caption{\sl \small
The contour plots of the lightest Higgs boson mass
for $\tan\beta = 10$ (left) and $\tan\beta = 20$ (right).
In both plots, we have taken a gluino mass of $1$\,TeV.
The green shaded region corresponds to $m_h>124$\,GeV
and the light-blue shaded region corresponds to $m_h>123$\,GeV.
The gray shaded region is excluded by tachyonic superparticles.
}
\label{fig:Higgs}
\end{center}
\end{figure}

\subsection{Relatively heavy lightest Higgs boson in Type-II model}
The ATLAS and CMS collaborations have reported interesting
hints of a Higgs boson with a mass around
$125$\,GeV\,\cite{ATLAS, CMS}.
In conventional gauge mediation models the $A$-terms are quite small. To get a lightest Higgs boson of this mass requires squark masses of $O(10)$\,TeV.
As we have seen, however,
a sizable $A$-term for the stop can be generated for $y_t' \simeq 1$
in the Type-II gauge mediation models. These large $A$-terms increase the mass of the lightest Higgs boson significantly \,\cite{Evans:2011bea}.

With a relatively large $A$-term the lightest Higgs boson mass,
which receives important SUSY breaking corrections from the top-stop loop
diagrams\,\cite{Okada,TU-363},
is pushed up to
\begin{eqnarray}
\label{eq:Higgs}
 m_{h^0}^2 \simeq m_Z^2 \cos^22\beta + \frac{3}{4 \pi^2}y_t^2 m_t^2 \sin^2{\beta}
\left( \log \frac{m_{\tilde t}^2 }{m_{t}^2 }+
\frac{A_t^2}{m_{\tilde t}^2}
-\frac{A_t^4}{12m_{\tilde t}^4}
\right)\ .
\end{eqnarray}
Here, $m_{Z}$ and $m_t$ are the masses of the $Z$-boson and top quark,
respectively.
The above expression for the Higgs mass is maximized for an $A$-term of order $A_t \simeq \sqrt{6} \times m_{\tilde t}$ (i.e. the hmax scenario).

In Fig.\,\ref{fig:Higgs}, we show a contour plot of the lightest Higgs boson mass
as a function of $y_t'$ and $F/M^2$ for $\tan\beta=10$ (left)
and $\tan\beta = 20$ (right).
To calculate the weak scale soft masses, we have used {\it SoftSusy}\,\cite{Allanach:2001kg},
and the lightest Higgs boson mass is calculated using {\it FeynHiggs}\,\cite{Frank:2006yh}.
In both panels, the green region corresponds to $m_h>124$\,GeV
and the light-blue shaded region corresponds to $m_h>123$\,GeV.
In our analysis, we used the central values of the top quark mass $m_t = 173.2\pm 0.9$\,GeV\,\cite{Lancaster:2011wr}
and the strong coupling constant, $\alpha_s(M_Z) = 0.1184\pm 0.0007$\,\cite{Bethke:2009jm}.
The gray shaded region in Fig.\,\ref{fig:Higgs}  for $x \simeq 1$ is excluded due to a very light stop, while the gray shaded region
for $y_{t'}\gtrsim 1$ and $x \ll 1$ is excluded due to a very light slepton.
Within the allowed region, we find that the vacuum stability condition\,\cite{Kusenko:1996jn},
\begin{eqnarray}
   A_t^2 + 3 \mu^2 < 7.5\, (m_{{\tilde t}_L}^2 + m_{{\tilde t}_R}^2 ) \ ,
\end{eqnarray}
is always satisfied.
Thus, the relatively large $A$-terms do not cause vacuum instability problems in Type-II models.

These figures shows that a relatively heavy Higgs boson is obtained for $y_t' \simeq 1$.
Notice that the lightest Higgs boson is heaviest in regions where $x\simeq 1$. Since $x\simeq 1$ corresponds to a low messenger scale, the suppression of the $A$-terms due to renormalization group evolution is less significant
in this region.
The figures also shows that the Higgs mass is only weakly dependent on $\tan\beta$ for much of the parameter space(see also
later discussion).

\begin{figure}[t]
\begin{center}
\begin{minipage}{.49\linewidth}
  \includegraphics[width=.9\linewidth]{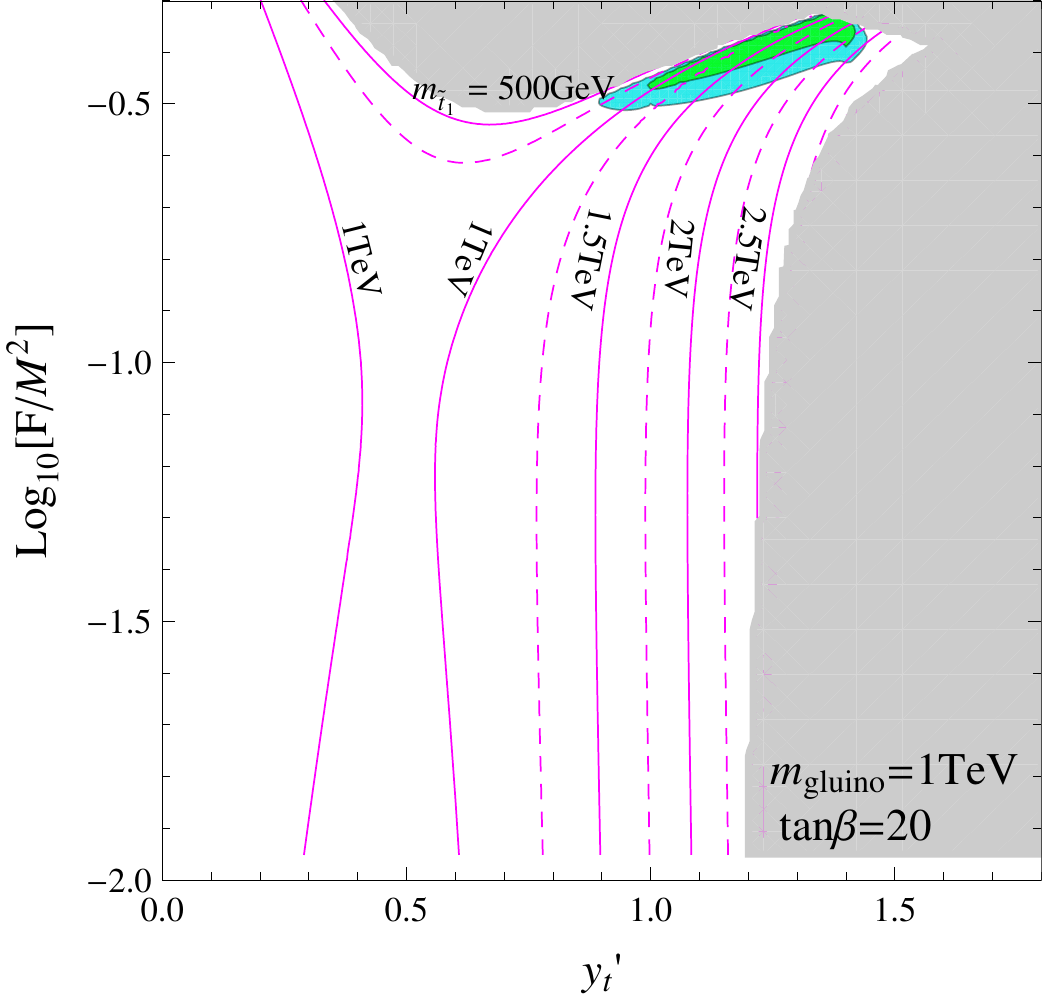}
  \end{minipage}
  \begin{minipage}{.49\linewidth}
  \includegraphics[width=.9\linewidth]{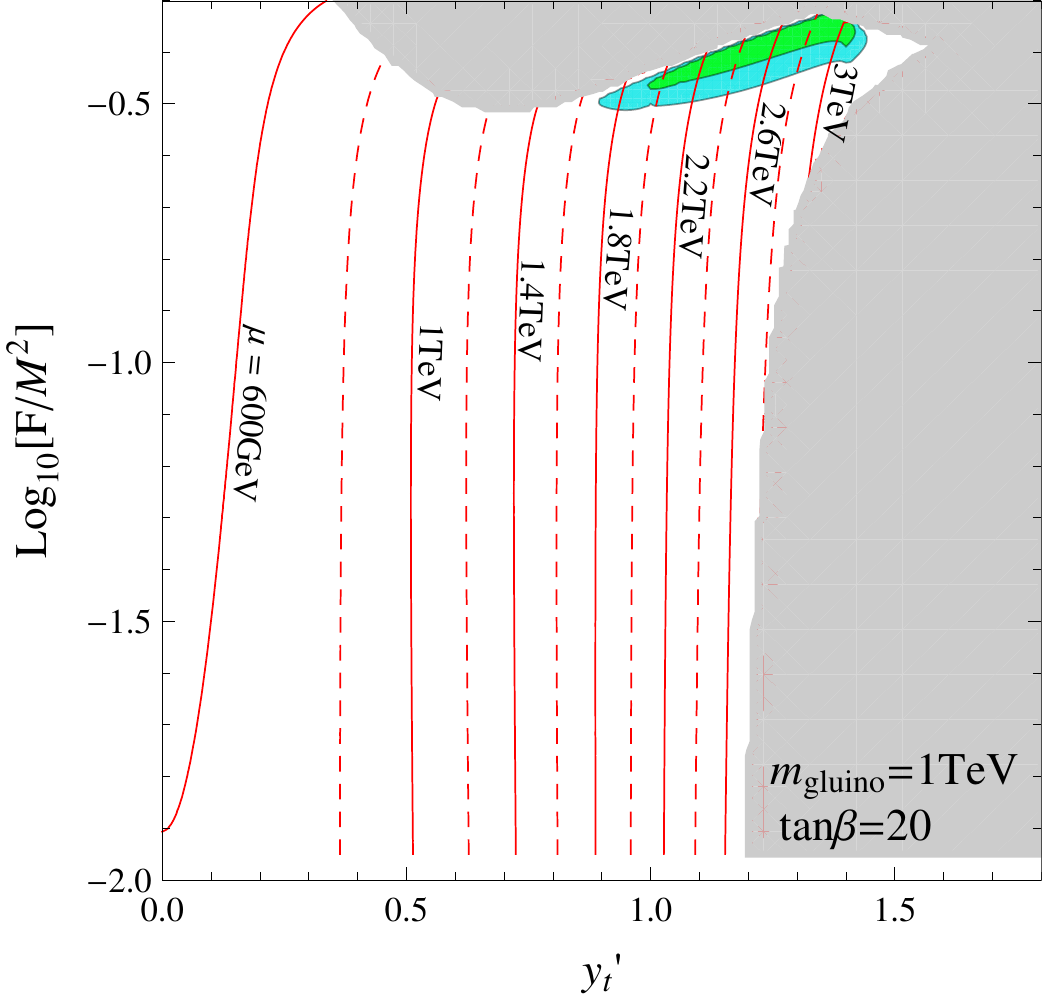}
  \end{minipage}
\caption{\sl \small
Contour plots of the lighter stop mass (left)
and the $\mu$-term for $\tan\beta = 20$ (right).
In both plots, we have taken a gluino mass of $1$\,TeV.
The results do not significantly depend on $\tan\beta$ as long as $\tan\beta \gtrsim 10$.
}
\label{fig:Stop}
\end{center}
\end{figure}

Before closing this section, we also show the lighter stop mass in Fig.\,\ref{fig:Stop}.
This figure shows how the stop mass increases with $y_t'$.
This increase in the stop masses is due to the two-loop contribution found in Eq.\,(\ref{eq:Twoloop}).
As a result, the stop becomes heavier for the bulk of the parameter space where the Higgs boson
mass is enhanced.
It should be also noted that the stops can be much lighter at the corner of the parameter region
for $x\simeq 1$,
where the negative one-loop contributions found in Eqs.\,(\ref{eq:oneloop1})
and (\ref{eq:oneloop2}) are important.
In particular, if the stop becomes significantly lighter than the gluino,
the gluino will decay mainly into a top and a stop,
which affects the search strategies for these models at the LHC.

In Fig.\,\ref{fig:Stop}, we also show contour plots of the $\mu$-term.
This figure shows that the $\mu$-term is relatively large even for the regions with a light stop.
This is due to the new two-loop contribution to $m_{H_u}^2$
given in Eq.\,(\ref{eq:Twoloop}). Because it is large and negative, a large $\mu$-term is needed to compensate.
As a result, the Higgsino masses are much heavier than the colored superparticles
in the region where the Higgs boson mass is largest.

\subsection{The muon anomalous magnetic moment}
Since the muon anomalous magnetic moment has been measure quite precisely,
it is an important probe of new physics beyond the Standard Model.
The current experimental value of the anomalous magnetic moment of the muon
is\,\cite{Bennett:2006fi}:
\begin{eqnarray}
a^{\rm exp}_\mu = 11659208.9(6.3)\times 10^{-10}\ .
\end{eqnarray}
The most recent calculation of the Standard Model prediction, on the other hand,
is \cite{Hagiwara:2011af};
\begin{eqnarray}
a_\mu^{\rm SM}= 11659182.8(4.9)\times 10^{-10}\ ,
\end{eqnarray}
which includes the updated data from  $e^+e^-\to $\,hadrons and the
latest evaluation of the hadronic light-by-light scattering contributions.
As a result, the experimental value of the muon $g-2$ significantly deviates from the Standard Model
prediction by about $3.3\sigma$, i.e.
\begin{eqnarray}
 \delta a_\mu = a_\mu^{\rm exp} - a_\mu^{\rm SM} = (26.1\pm 8.0)\times 10^{-10}\ .
\end{eqnarray}

It is quite tantalizing that this deviation can be explained by the existence of superparticles. The supersymmetric contribution to the anomalous magnetic moment of the muon is proportional
to $\tan\beta$ and is suppressed for the heavy superparticle masses.
For a precise expression of the supersymmetric contribution to the muon
$g-2$, see Ref.\,\cite{Cho:2011rk}.
Unfortunately, however,
the relatively heavy Higgs boson hinted at by ATLAS and CMS collaborations
requires rather heavy superparticle masses in most models.
Therefore, it is not easy to realize both a Higgs boson mass of around $125$\,GeV
and a consistent muon $g-2$ simultaneously.

The above tension between a relatively heavy Higgs boson mass
and a sizable supersymmetric contribution to the muon $g-2$
is eased in Type-II models.
As we have seen, the relatively heavy lightest Higgs boson
can be realized even when the other superparticles are relatively light.
This feature is quite advantageous for simultaneously explaining both the heavy Higgs boson mass
and the deviation of the muon $g-2$.

Another advantage Type-II models have is light left-handed sleptons.
The rather light left-handed sleptons are due to renormalization group evolution,
\begin{eqnarray}
 \frac{d}{dt} m_{\rm slepton}^2 = -\sum_{a=1,2}8C_a\frac{g^2_a}{16\pi^2} |M_{a}|^2
 +\frac{1}{8\pi^2}\frac{3}{5} Y g_1^2 {\cal S}\ ,
\end{eqnarray}
where $M_a$ denote the gaugino masses,
$C_{2} = 3/4$ and $Y = - 1/2$  for the doublet sleptons,
and
$C_2 = 0$ and $Y = 1$ for the right handed sleptons.
$\cal S$ is given by,
\begin{eqnarray}
{\cal S} &=& \tr\left[ Y_{i} m_{i}^2 \right] =
m_{H_u}^2 - m_{H_d}^2
+ \tr\left[
m_{Q}^2
-m_{L}^2
-2m_{\bar{U}}^2
+m_{\bar{D}}^2
+m_{\bar{E}}^2
\right]\ .
\end{eqnarray}
The purely gauge mediated contributions to the above expression cancel at the messenger scale. However, the new contributions to the soft masses that are proportional to $y_t'$ do not cancel.%
\footnote{This means that the spectrum of Type-II gauge mediation
deviates from the prediction in General Gauge Mediation\,\cite{Meade:2008wd}.}
As we see from  Eq.\,(\ref{eq:Twoloop}),
the two-loop contribution to $m_{Q_3, \bar{T}}^2$
are  large and  positive for $y_t' \gtrsim 1$ which leads
to a negative ${\cal S}$.
The negative tree-level contribution to $m_{H_d}^2$ for $x \simeq 1$ also gives a negative contribution to
${\cal S}$.
Therefore, through renormalization group running,
the doublet sleptons become lighter at the low energy scale,
while the right-handed sleptons become heavier.%
\footnote{
The squark masses also receive a similar, but less significant, renormalization group effect
from ${\cal S}$ with the signs depending on their $U(1)$ hypercharges.
}
This suppression of the left handed slepton mass is also important for
obtaining a sizable supersymmetric contribution to the muon $g-2$ in Type-II model.

\begin{figure}[t]
\begin{center}
\begin{minipage}{.49\linewidth}
  \includegraphics[width=.9\linewidth]{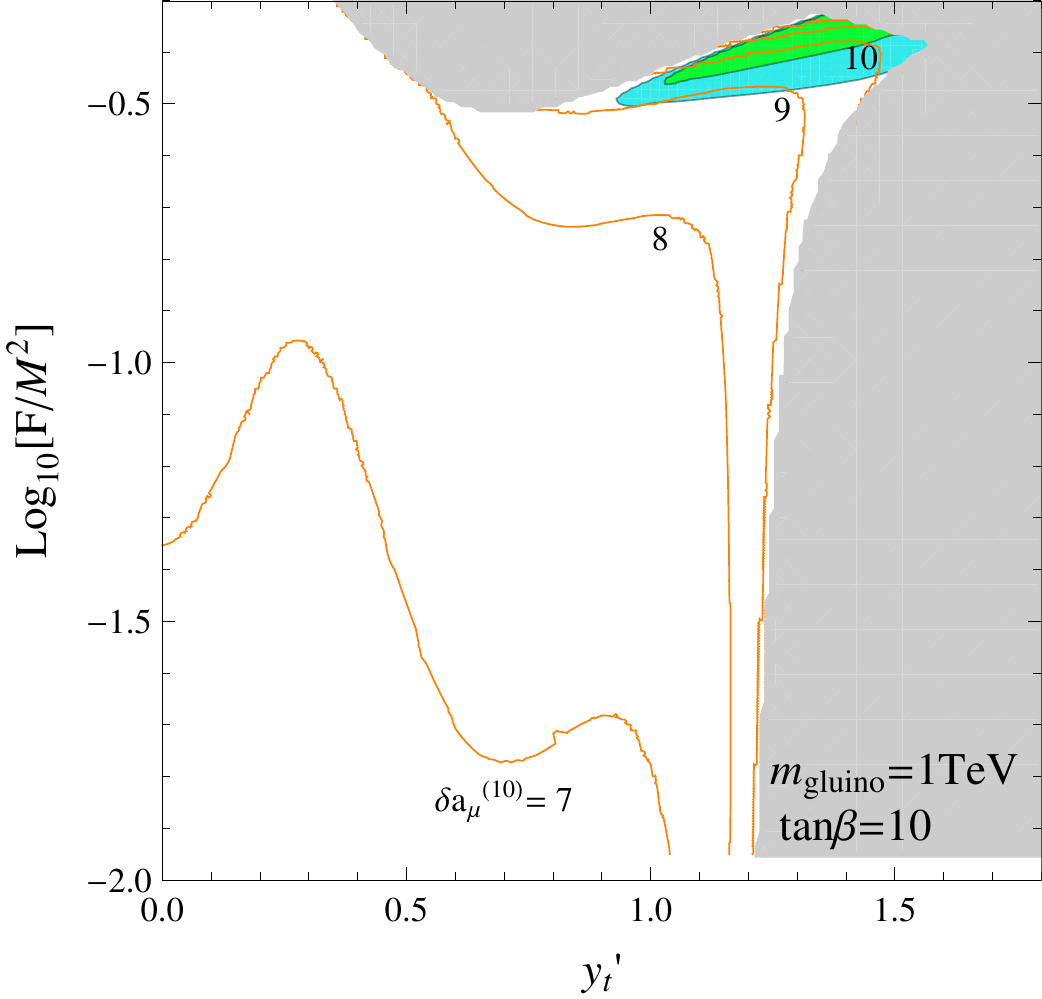}
  \end{minipage}
  \begin{minipage}{.49\linewidth}
  \includegraphics[width=.9\linewidth]{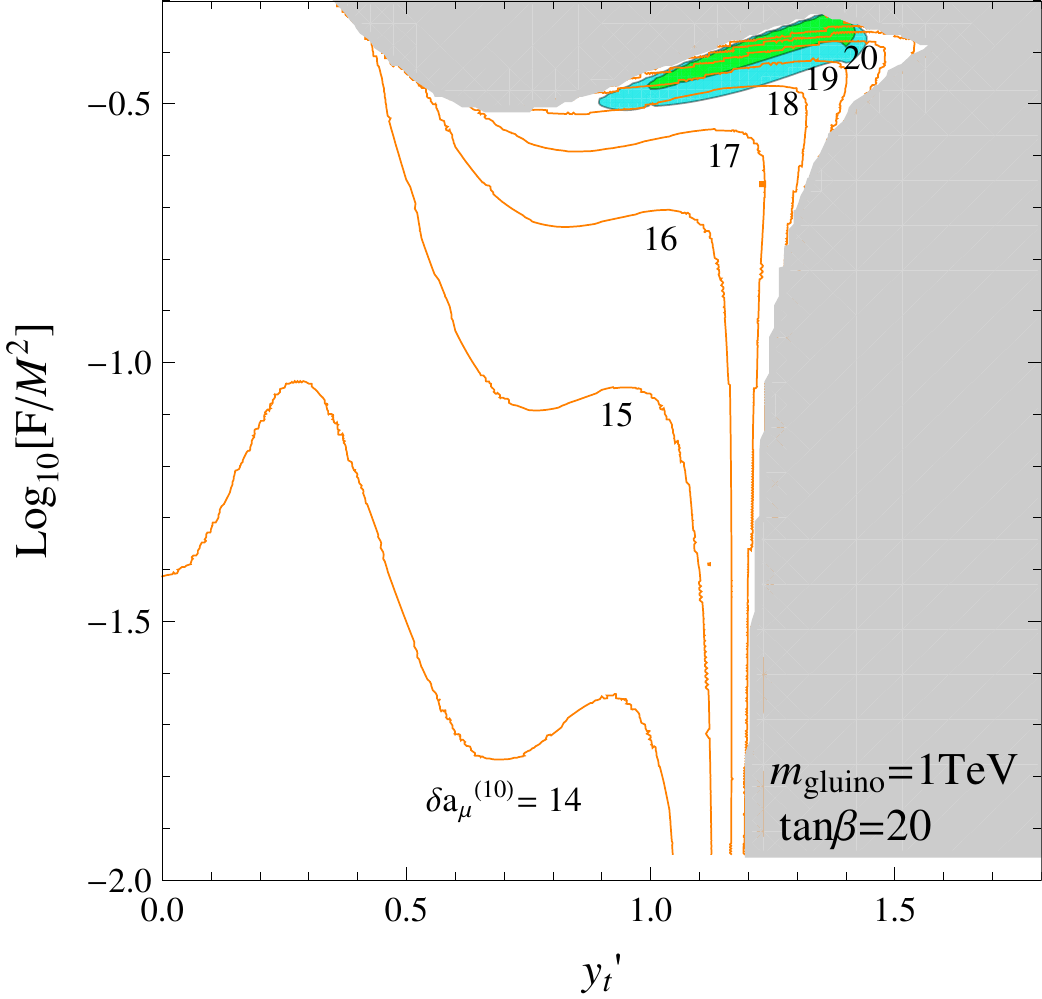}
  \end{minipage}
\caption{\sl \small
The contour plots of the supersymmetric contribution
to the muon anomalous magnetic moment $\delta a_{\mu}^{\rm SUSY}\times 10^{10}$
for  $\tan\beta = 10$ (left) and $\tan\beta = 20$ (right).
The green and light-blue shaded regions are same as in Fig.\,\ref{fig:Higgs}.
}
\label{fig:gmuon}
\end{center}
\end{figure}

\begin{figure}[t]
\begin{center}
\begin{minipage}{.49\linewidth}
  \includegraphics[width=.9\linewidth]{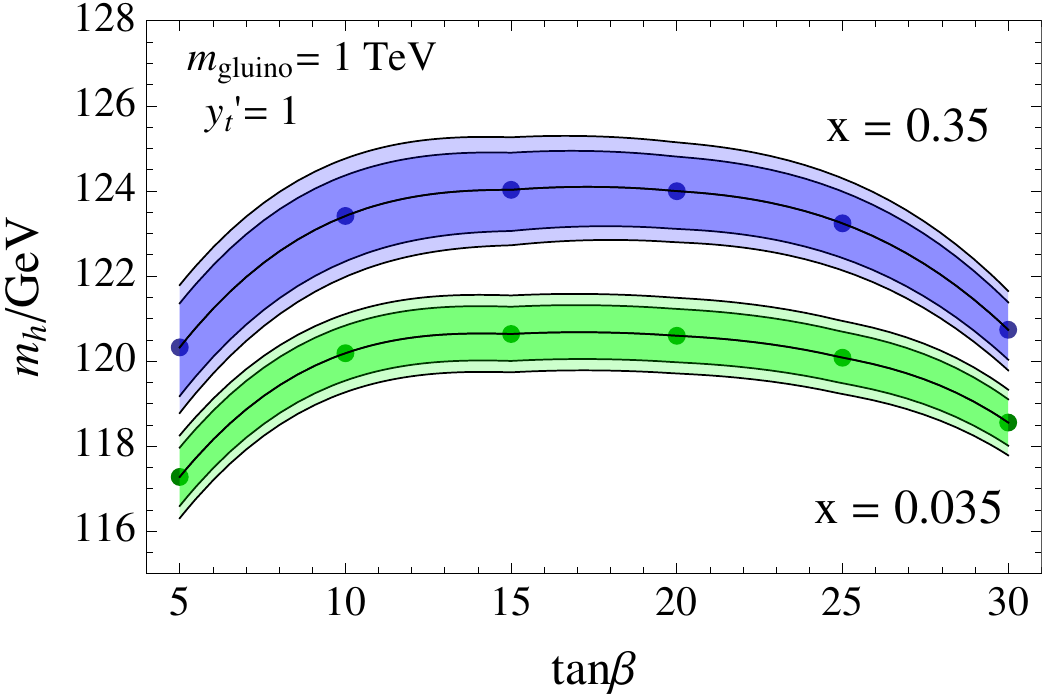}
  \end{minipage}
  \begin{minipage}{.49\linewidth}
  \includegraphics[width=.9\linewidth]{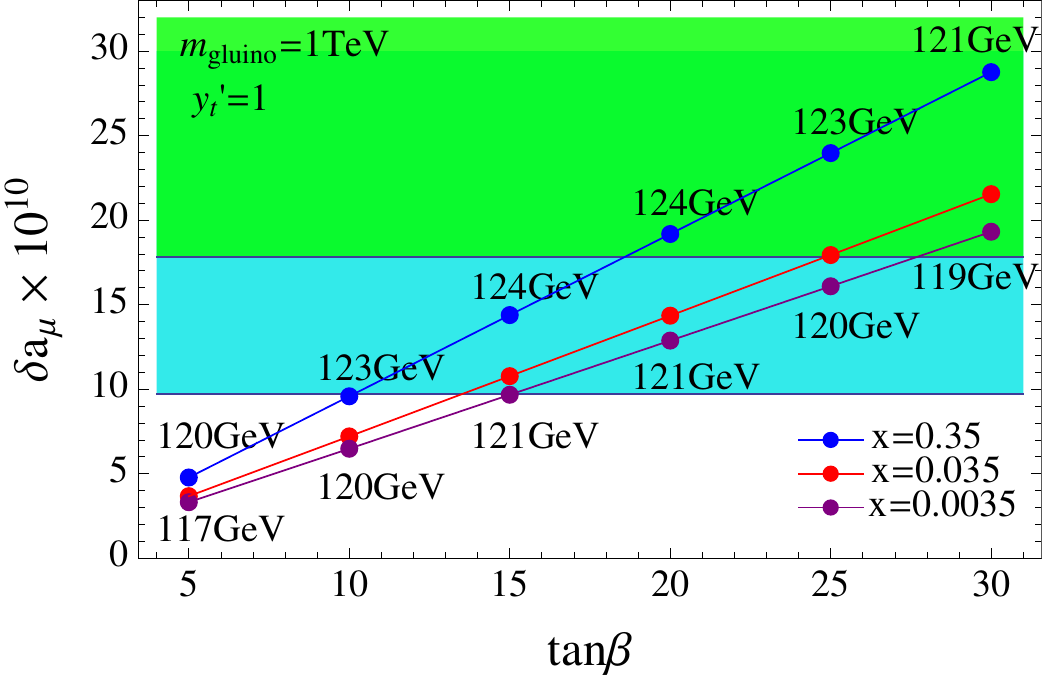}
  \end{minipage}
\caption{\sl \small
(Left) The $\tan\beta$ dependence of the lightest Higgs boson mass for $m_{\rm gluino}=1\,$TeV.
(Right) The $\tan\beta$ dependence of the muon g-2  for $m_{\rm gluino}=1\,$TeV.
The green shaded region corresponds to the muon $g-2$ consistent with
the experimental value at the $1\sigma$ level.
We also show the lightest Higgs boson mass for each parameter point.
The lightest Higgs boson masses are degenerate for $x = 0.035$ and $x = 0.0035$.
}
\label{fig:tanb}
\end{center}
\end{figure}

In Fig.\,\ref{fig:gmuon}, we show contour plots of the supersymmetric contribution
to the muon $g-2$
as a function of $y_t'$ and $F/M^2$ for both $\tan\beta=10$ (left) and $\tan\beta = 20$ (right).
Here, we have taken $\mu>0$ so that the supersymmetric contribution
shifts the muon $g-2$ in the right direction.
The muon $g-2$ is calculated using {\it FeynHiggs}.
By comparing the results for $\tan\b =10$ and $\tan\b = 20$, we see that
$\delta a^{\rm SUSY}_{\mu}$ is proportional to $\tan \b$ as expected.
This figure shows that a muon $g-2$ consistent with the experimental value
at the $1\sigma$ level can be realized for $\tan\beta = 20$.

In Fig.\,\ref{fig:tanb}, we show the $\tan\beta$ dependence of the lightest Higgs boson mass
and muon $g-2$ for $m_{\rm gluino} = 1$\,TeV and $y_t' = 1$.
The bands on the Higgs boson mass show the
uncertainties of the Higgs mass which were estimated by {\it FeynHiggs}.
We also find that the $1\sigma$ error on the top quark mass $m_{\rm top}= 173.2\pm 0.9$\,GeV
lead to similar uncertainties in the lightest Higgs boson mass which we have not shown here.
For $x<0.035$,  the lightest Higgs boson mass does not change
if the other parameters are fixed(see also Fig.\,\ref{fig:Higgs}).
This figure shows that the lightest Higgs boson mass is saturated for $\tan\beta \simeq 15-20$.
The muon $g-2$, on the other hand, is proportional to $\tan\beta$.
Therefore,
we find that $\tan\beta \simeq 20$ and $x \simeq 0.3$ is most advantageous
for simultaneously explaining a relatively Heavy Higgs boson and the deviation in the muon $g-2$.

\begin{figure}[t]
\begin{center}
\begin{minipage}{.49\linewidth}
  \includegraphics[width=.9\linewidth]{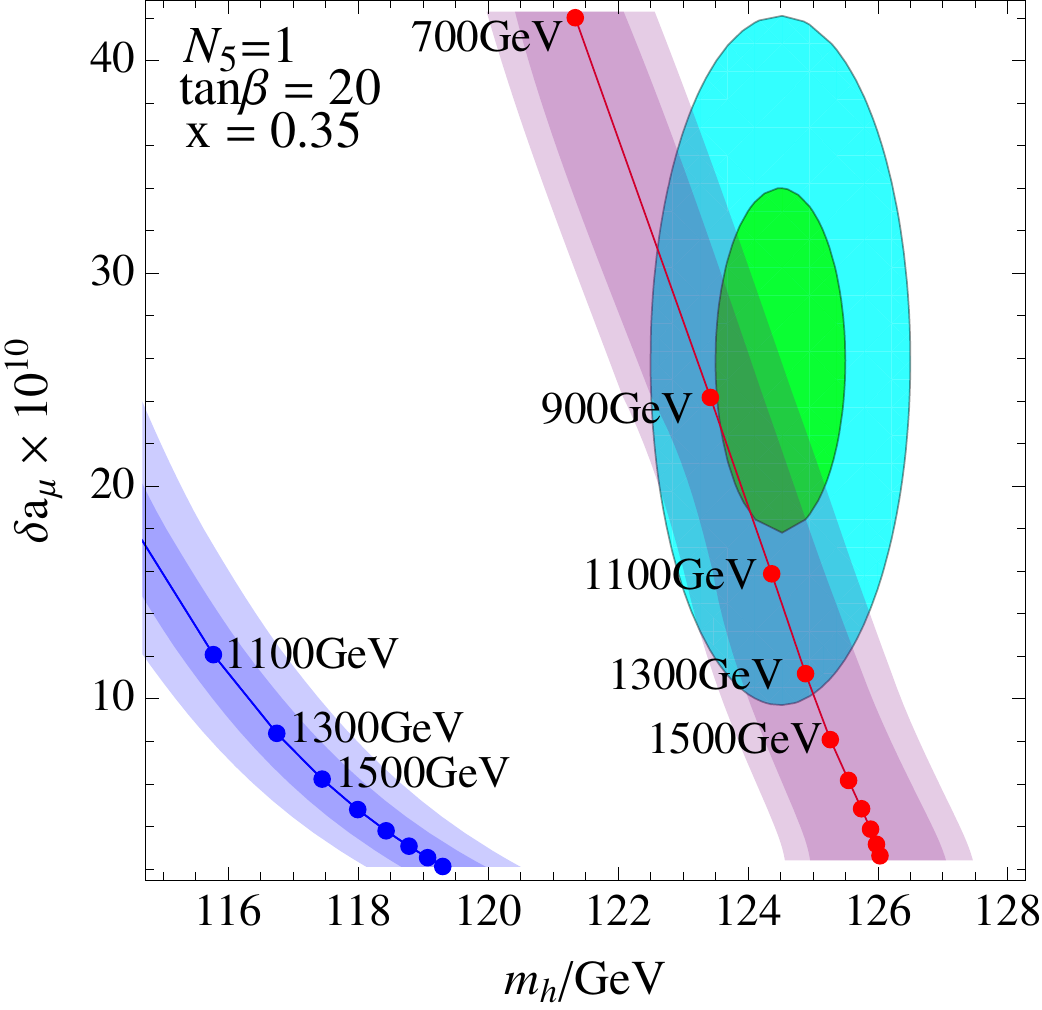}
  \end{minipage}
  \begin{minipage}{.49\linewidth}
  \includegraphics[width=.9\linewidth]{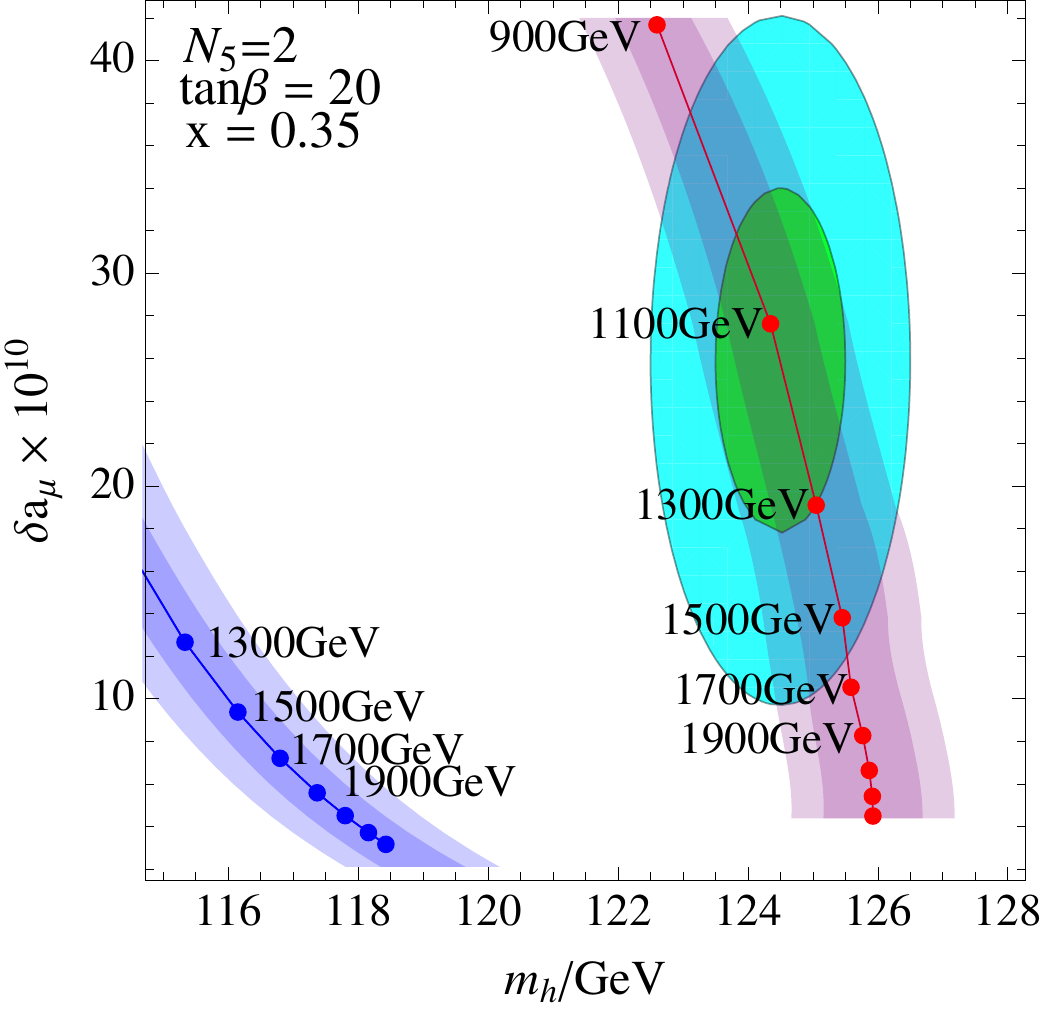}
  \end{minipage}
\caption{\sl \small
The correlation between the lightest Higgs boson mass and the muon  $g-2$
for $y_t' = 1$ (red) and $y_t' = 0$ (blue).
Each band corresponds to uncertainties of the Higgs mass estimated by FeynHiggs.
The definition of the oval region is given in the text.
The gluino masses for a each point are also shown.
}
\label{fig:N5}
\end{center}
\end{figure}

In Fig.\,\ref{fig:N5}, we show the predicted value of the lightest Higgs boson
mass and the muon $g-2$ for $y_t'  = 1$ (red) and $y_t' = 0$ (blue).
We have fixed $\tan\beta  = 20$ and $x = 0.35$.
The oval regions correspond to  $\chi<1$ (green) and $\chi<2$ (blue), respectively,
where $\chi$ is defined by,
\begin{eqnarray}
\label{eq:oval}
\chi = \left(
\frac{(m_h - 125\,{\rm GeV})^2}{\sigma_h^2}
+
\frac{(\delta a_\mu^{\rm SUSY} - 26.1 \times 10^{-10})^2}{(8.0\times 10^{-10})^2}
\right)^{1/2}\ .
\end{eqnarray}
Here, we have used $\sigma_h = 1\,$GeV for illustrative purpose.
This figure shows that the relatively heavy Higgs boson mass and a consistent muon $g-2$
at the $1\sigma$ level can be realized simultaneously for $m_{\rm gluino} \simeq 1$\,TeV.

So far, we have considered a minimal messenger sector, i.e. $N_5 = 1$.
For $N_5 > 1$, the gauginos become relatively heavier for the same squark/slepton masses.
Thus,  the muon $g-2$ can be explained for a relatively heavier gluino mass
compared to the models with $N_5 = 1$.
In the right panel of Fig.\,\ref{fig:N5}, we show the lightest Higgs boson
mass and the muon $g-2$ for $N_5= 2$.
As expected, this figure shows that muon $g-2$ can be explained even for a heavier gluino mass.
As has been mentioned above, the left handed sleptons are light in the regions where the lightest Higgs mass is enhanced.
Thus, we find that the next lightest superparticles (NLSP)
is a stau if $N_5 > 1$.
As we will see in the next section, a stable stau NLSP scenario can be
easily excluded by the LHC experiments in the near future if
we require the muon $g-2$ be consistent with the experimental value at the $1\sigma$ level.

\subsection{Perturbativity of $\tilde{y}_U$}
In the above analysis, we have taken $y_t' = 1$ as a bench mark point
to explain both the Higgs mass around $125$\,GeV
and the observed muon $g-2$.
Such a relatively large Yukawa coupling constant, however, often has a Landau pole
below the GUT scale, ruining one of the important
motivation for the SSM.
In this subsection, we discuss the constraints on the Yukawa coupling constant

Type-II models above the messenger scale are well described by the superpotential given in Eq.\,(\ref{eq:superpotential1}).
Therefor, in the high energy theory we need to consider the perturbativity of $\tilde{y}_t$. This coupling is related to our low-scale parameters at the messenger scale through the expression 
\begin{eqnarray}
\tilde{y}_t = (y_t^2 + y_t'^2 )^{1/2} \ 
\end{eqnarray}
Therefore, the  perturbativity constraint on $\tilde{y}_t$ is more stringent
than the usual constraint on $y_t$ in the MSSM, since $\tilde{y}_t$ is larger than
$y_t$ at the messenger scale.

\begin{figure}[t]
\begin{center}
\begin{minipage}{.49\linewidth}
  \includegraphics[width=.9\linewidth]{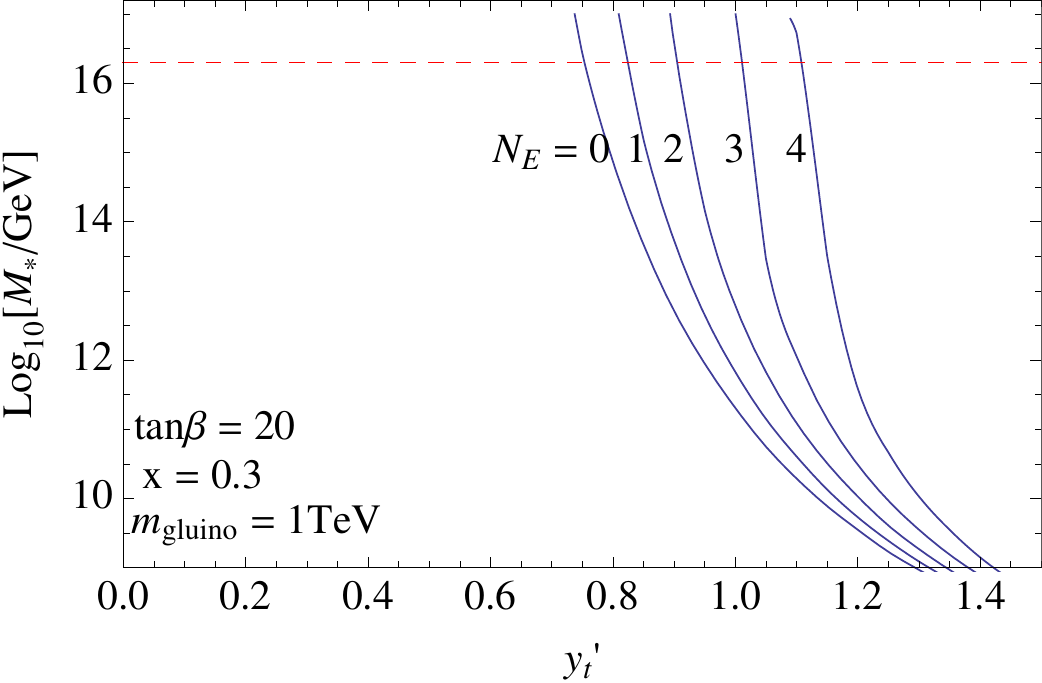}
  \end{minipage}
  \begin{minipage}{.49\linewidth}
  \includegraphics[width=.9\linewidth]{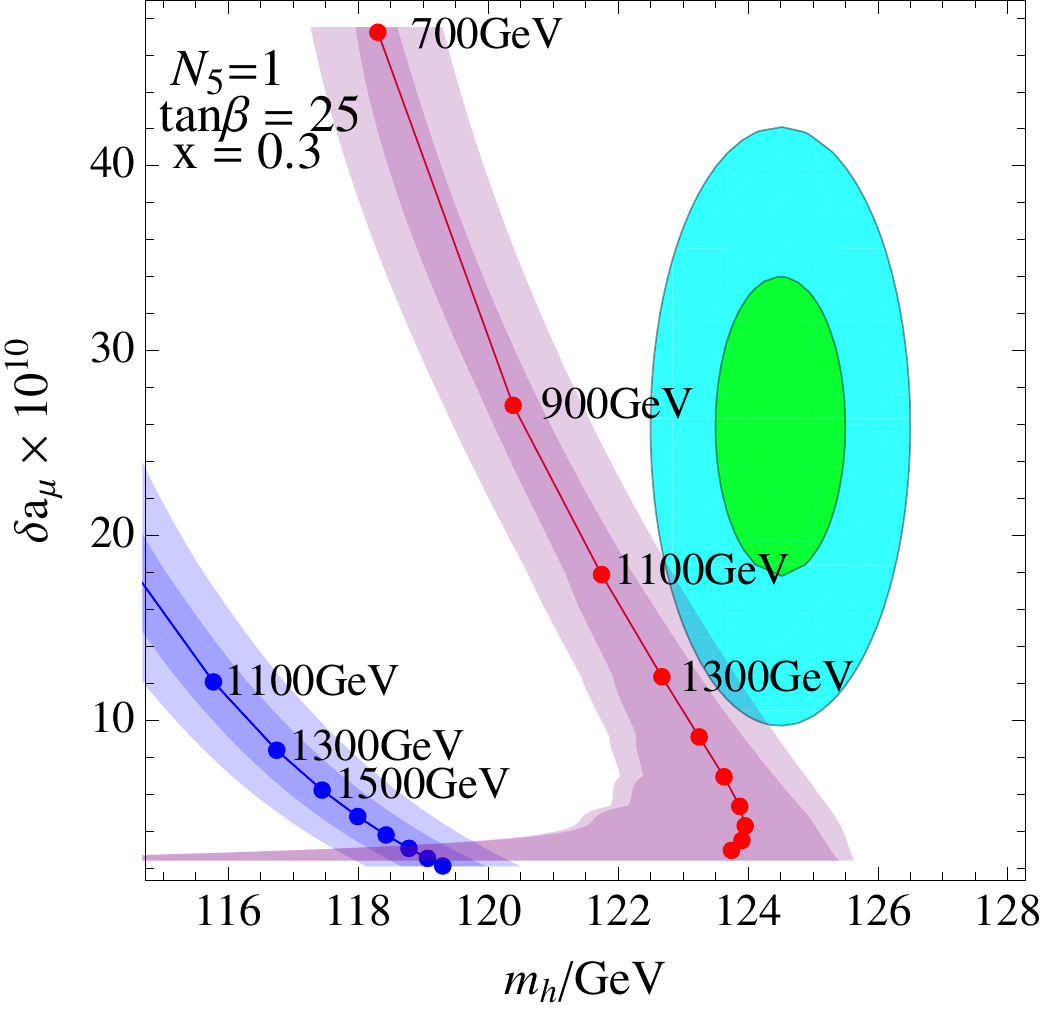}
  \end{minipage}
\caption{\sl \small
(Left) The scale $M_*$ where $\tilde{y}_t$ becomes non-perturbative
for $N_E = 0-4$. For $N_E > 0$, we assumed that
the mass of the extra vector-like multiplet is $1$\,TeV.
The dashed line denotes the GUT scale about $2\times 10^{16}$\,GeV.
(Right)
The correlation between the lightest Higgs boson mass and the muon  $g-2$
for $y_t' = 0.75$ (red) where the $\tilde{y}_t$ is perturbative up to the GUT scale.
}
\label{fig:YTP}
\end{center}
\end{figure}

Assuming $g, g' \ll O(1)$, the renormalization group equation
of $\tilde{y}_t$ is identical to that of $y_t$ which is given by,
\begin{eqnarray}
\label{eq:RGEYT}
\frac{d}{d t}\tilde{y}_{t}  = \frac{\tilde{y}_t}{16\pi^2} \left(
6 \tilde{y}_t^2 + y_{b}^2 + y_\tau^2
-\frac{16}{3}g_3^2
-3 g_2^2
-\frac{13}{15}g_1^2
\right) \ ,
\end{eqnarray}
and all other parameters appearing here are the standard couplings of the MSSM.
In Fig.\,\ref{fig:YTP}, we show the scale $M_*$ for which $y_t'$ becomes non-perturbative,
i.e. $y_t' (M_*)\simeq 4\pi$.
Notice that the scale $M_*$ is not sensitive to $\tan\beta$ or deviations in $m_{\rm gluino} = O(1)$\,TeV.

Fig.\,\ref{fig:YTP} shows $y_t'$ needs to be $\lesssim 0.75$ for the theory to be perturbative up to the GUT scale.
Thus, the bench mark point we have taken in the previous section is plagued by a
Landau pole below the GUT scale.
In the right panel of Fig.\,\ref{fig:YTP}, we replot Fig.\,\ref{fig:N5} but take $y_t' = 0.75$.
The figure shows that even in this case, the muon $g-2$ can be
within $1\sigma$ of the experimental value for $m_h\simeq123$\,GeV
and be within $2\sigma$ for $m_h \simeq 125$\,GeV.

 One simple way to make the theory perturbative up to the GUT scale for $\tilde{y}_t=1$ is to make the MSSM gauge interactions more asymptotic non-free.%
 \footnote{Another simple way to ameliorate the Landau problem is
to introduce additional $U(1)$  gauge symmetry. This choice would alter the beta function of $\tilde{y}_t$ making it more asymptotically free.
 }
 The larger gauge coupling constants
 at the higher energy suppress $\tilde{y}_t$ at the higher energy via the renormalization
 group equation in Eq.\,(\ref{eq:RGEYT}).
 In Fig.\,\ref{fig:YTP}, we show the scale $M_*$ in the presence of $N_E$ extra matter multiplets
each with a mass around $1$\,TeV.%
\footnote{Here, $N_E = 1$ corresponds to an extra matter multiplet of ${\mathbf 5}+{\mathbf 5}^*$.}
The figure shows that three pairs of ${\mathbf 5}+{\mathbf 5}^*$
(or a pair of ${\mathbf 10}+{\mathbf 10}^*$) is enough
to make the bench mark point $y_t' = 1$ perturbative up to the GUT scale\footnote{The bench mark point with $y_t'=1$ and no extra matter multiplets ($N_E=0$) has a Landau pole at around $10^{11}$\,GeV. This may be regarded as an indication of a rather intriguing posibility that the Higgs and top quarks are composite states of some new strongly interacting theory at $10^{11}$\,GeV \cite{Randall, BPY}}.

\section{Constraints and Prospects at the LHC}\label{sec:LHC}
In this section, we discuss the constraints and prospect of discovery for the present model at the LHC.
At the LHC, the production cross section of SUSY particles is tightly linked to the first family squark masses and gaugino masses.
As we have discussed, the stops are rather heavier in the bulk of the parameter region
which realizes a rather heavy Higgs particle $m_h\simeq 125$\,GeV (see Fig.\,\ref{fig:Stop}).
Thus, the LHC signature of our model is very similar to the usual minimal gauge mediation models
for most cases.

As was shown above, $m_{\rm gluino} \simeq 1$\,TeV and $x ={\cal O}(0.1)$
are best suited to explain the Higgs mass and the deviations in the muon $g-2$.
Gauge mediation with this messenger scale can be tested at the LHC rather easily.
The LHC signatures, however, are strongly dependent on the gravitino mass.
As we discussed above, the gravitino mass tends to be not very light in
Type-II gauge mediation.
Thus, we assume that the NLSP is stable inside the detectors of the LHC experiments.
In the present model, the possible NLSP are the lightest neutralino ($\tilde{\chi}^0_1$) or
the stau ($\tilde{\tau}_1$).
In the former case, the key signature is high $p_{\rm T}$ jets plus large missing energy.
This signature is a very similar signature to conventional gravity mediation models.
In the latter case, the stau penetrates the detectors and heavy charged tracks would be observed.

\begin{figure}
\subfloat[{\it \small $y_t'$ vs $1/x$ for $|a_{\mu} - a_{\mu}^{\rm exp}|<1 \sigma$}. ]{
{\includegraphics[clip, width=0.5\columnwidth]{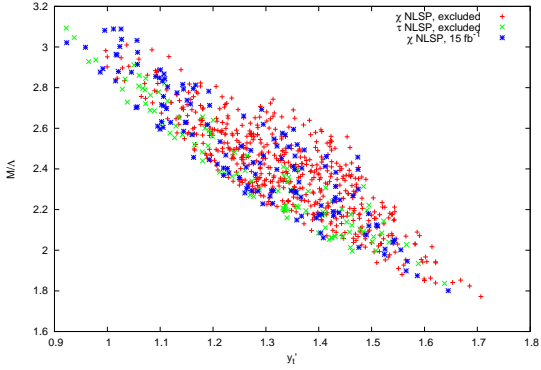}}
}
\subfloat[{\it\small  $m_{\rm stop}$ vs $m_{\rm gluino}$ for $|a_{\mu} - a_{\mu}^{\rm exp}|<1 \sigma$.}]{
{\includegraphics[clip, width=0.5\columnwidth]{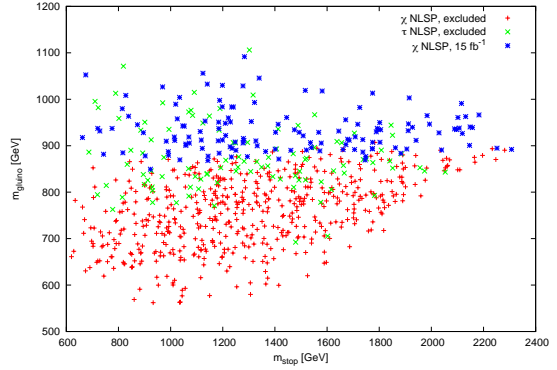}}
}\\

\subfloat[{\it \small $y_t' $ vs $1/x$ for  $1 \sigma<|a_{\mu} - a_{\mu}^{\rm exp}|<2 \sigma$.}]{
{\includegraphics[clip, width=0.5\columnwidth]{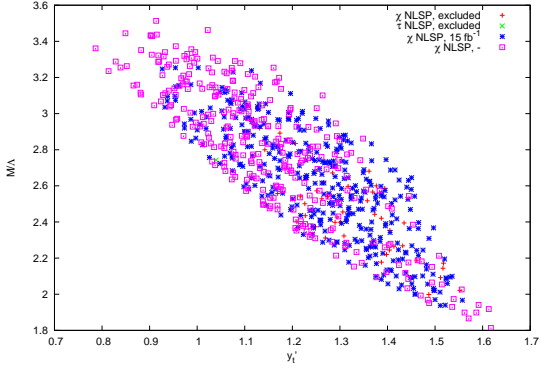}}
}
\subfloat[ {\it \small $m_{\rm stop}$ vs $m_{\rm gluino}$ for $1 \sigma<|a_{\mu} - a_{\mu}^{\rm exp}|<2 \sigma$.}]{
{\includegraphics[clip, width=0.5\columnwidth]{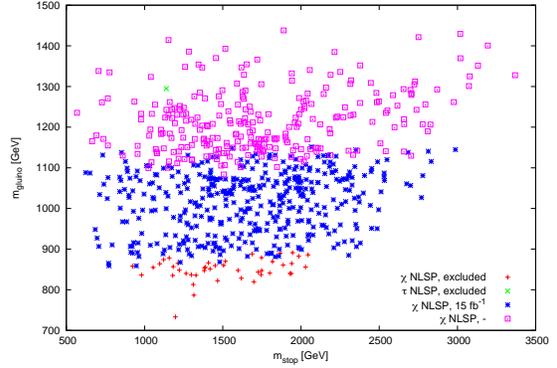}}
}
\caption{Scatter plots of the model parameters and the MSSM soft masses
showing collider constraints and prospects for each point.
Red crosses are the regions where the neutralino is the NLSP and has already been excluded
at the LHC, green crosses have the stau NLSP and are also already excluded,
blue stars have a neutralino NLSP and will be excluded by the $7$\,TeV run at the LHC,
and purple squares are neutralino NLSP and cannot be excluded at LHC with a $7$\,TeV
center of mass energy.
The sup mass is well fitted by $m_{\rm sup} \simeq 1.25\,m_{\rm {gluino}}$.
}
\label{fig:scatter}
\end{figure}

In Figs.\,\ref{fig:scatter}, we show the results of a parameter scan over the in puts of the present model.
We also require $m_{h}>123.5$ GeV and $a_\mu$ within $1$-$2\sigma$ of the experimental value.
To estimate the LHC constraints, we have used the data of 
Refs\,.\cite{Aad:2011ib,ATLAS:2011ad,CHAMP} and
the program
ISAJET 7.72 \cite{Paige:2003mg} to generate the MSSM mass spectrum and decay tables
and Herwig 6.510 \cite{Marchesini:1991ch,Corcella:2002jc,Corcella:2000bw,Corcella:2002jc} to generate SUSY events at the LHC.
For the detector simulation, we have used AcerDet\,1.0\cite{RichterWas:2002ch} which was slightly modified by the authors.
We also show the prospects of excluding these models for higher integrated luminosity, 15\,fb${}^{-1}$.
We have used the following cuts, 4-jets mode:
 at least four jets, with  $p_{\rm T}>100$\,GeV.  $E_{\rm T}^{\rm miss}>200$\,GeV,
 $M_{\rm eff}>1200$\,GeV and $E_{\rm T}^{\rm miss}/M_{\rm eff}>0.3$, or a 2-jets mode
 with the leading jet having $p_{\rm T}>300$\,GeV, and the other jet having $p_{\rm T}>200$\,GeV, and $E_{\rm T}^{\rm miss}>400$\,GeV,
 $M_{\rm eff}>1200$\,GeV and $E_{\rm T}^{\rm miss}/M_{\rm eff}>0.35$.
 To estimate the Standard model background,
 we have used the programs MC@NLO 3.42\,\cite{Frixione:2002ik}
(for $t\bar{t},WW,WZ$ and $ZZ$),
Alpgen 2.13\,\cite{Mangano:2002ea} (for $Wj,Zj$ and $W/Z+b\bar{b}/t\bar{t}$).

As pointed out in Ref.\,\cite{CHAMP}, the long-lived stau mass is strongly constrained, $m_{\tilde{\tau}_1}\gtrsim 290$\,GeV.
This constraint is inconsistent with a large $\delta a_{\mu}^{\rm SUSY}$ for the points
in the stau NLSP region.
For larger values of $N_5$, the stau tends to be the NLSP. This makes it difficult to find models that are consistent with both the muon $g-2$ measurements and LHC collider phenomenally.

In the case of a neutralino NLSP, the LHC is less constraining and there are regions that are consistent with a large $\delta a_{\mu}^{\rm SUSY}$. The present constraints on the gluino mass are
 $m_{\rm {gluino}} \gsim 900$\,GeV. After the complete $\sqrt{s} = 7$\,TeV LHC run,
 the constraints on the gluino mass could be pushed to  $m_{\tilde{g}} \gsim 1100$\,GeV.
 Once the LHC is upgraded to $\sqrt{s} = 8$\,TeV,
 its reach could be extend to $m_{\tilde{g}} \sim 1200$\,GeV.
Requiring  $a_\mu$ within 1$\sigma$, we can see that almost all regions can be tested at the LHC with an integrated luminosity of 15\,fb${}^{-1}$.
For $1 \sigma<|a_{\mu} - a_{\mu}^{\rm exp}|<2 \sigma$, the soft mass can be larger
by about a factor of $1.4$.
Because of the larger superparticle masses, it is unlikely that the $7$\,TeV or $8$\,TeV LHC run will be capable of detecting the superparticles for this parameter space.
However, the 14 TeV LHC can still easily exclude this region or possibly make a discover.

\section{Conclusions and Discussions}
In this paper, we have revisited the lightest Higgs boson mass in Tpye-II gauge mediation.
We have shown that a Higgs boson mass around $125$\,GeV can be realized
even for a gluino mass as light $m_{\rm gluino}\sim 1$\,TeV.
Interestingly, we have also found that the muon anomalous magnetic moment can be
consistent with the experimental value, at the $1\sigma$ level, even for this relatively heavy Higgs boson.
It was also shown that much of the parameter space can be checked
at the LHC experiments in the near future.

We emphasize again that the field content of this more generic gauge mediation
is the same as minimal gauge mediation, and the
only real difference is the newly added interaction
Eq.(\ref{eq:top}).
It is surprising that such a small extension of minimal gauge mediation
can resolve many of its difficulties, i.e. a relatively heavy Higgs
boson and the deviation of the muon $g-2$.

Finally, let us comment on possible dark matter candidates for these models.
As we mentioned above, the gravitino is expected to be heavier.
Thus, it can be the dark matter candidate if the reheat temperature is appropriately chosen\,\cite{Moroi:1993mb}.
Furthermore, a $1$\,GeV gravitino dark matter can be consistent with thermal leptogenesis\,\cite{leptogenesis}.%
\footnote{The detailed analysis of the gravitino dark matter including the
thermal history of the messenger/supersymmetry breaking sectors will
be discussed elsewhere.
We also mention that
the gravitino dark matter scenario with mass lighter than $1$\,GeV
can also be consistent with thermal leptogenesis
if there is sufficient entropy production\,\cite{Fujii:2003iw,Ibe:2010ym}.
}

\appendix
\section{
Realization of Type-II Gauge Mediation
}\label{sec:symmetry}
In this appendix, we give the symmetries which realize the Type-II gauge mediation.
In Table\,\ref{tab:charge}, we show the $R$-symmetries and $U(1)$
symmetry, which is only by the positively charged spurion $\phi_+$.
Under these charge assignments, the generic superpotential at the renormalizable level
is given by,
\begin{eqnarray}
W = Z \tilde{\Phi} \bar{\Phi} + \phi_+ Z \tilde{H}_u \bar{\Phi} + Z^3 + \tilde{\mu} \tilde{H}_u H_d +
\mbox{(MSSM Yukawa interactions)}\ .
\end{eqnarray}
Here, we have omitted the coupling constants for convenience.
The cubic term for $Z$ is important for cascade supersymmetry
breaking\,\cite{Ibe:2010jb,Evans:2011pz},
although, the Type-II mechanism can be applied
to other models which have supersymmetry breaking
spurions with both an $A$-term and $F$-term expectation value.

For example, the unwanted term
\begin{eqnarray}
W =  \tilde{\Phi}\, {\mathbf 10}\,{\mathbf 10} \ ,
\end{eqnarray}
which could cause FCNC or rapid proton decay is forbidden by
the holomorphic property of the superpotential, i.e. the SUSY zero mechanism.
The other problematic term,
\begin{eqnarray}
W = \bar{ \Phi}\, {\mathbf 10}\,{\mathbf 5}^* \ ,
\end{eqnarray}
is forbidden by the $R$-symmetry.
Furthermore, the above symmetries also forbid unnecessary mass terms,
\begin{eqnarray}
W= \tilde{\Phi}_{\bar{L}}\, H_d\ +  \bar{\Phi}_{\bar{L}}\, H_u\ .
\end{eqnarray}

\begin{table}[t]
\caption{\sl\small
The charge assignments for the broken $U(1)$ symmetry are presented here.
We have used $SU(5)$ GUT representations for the MSSM matter fields,
i.e. ${\mathbf 10} = (Q_L,\bar{U}_R,\bar{E}_R) $
and ${\mathbf 5}^* = (\bar{D}_R,L_L) $.
We also show the charge of the right-handed neutrinos $\bar{N}_R$, which is need for the see-saw mechanism\,\cite{seesaw}.
}
\begin{center}
\begin{tabular}{|c|c|c|c|c|c|c|c|c|c|}
\hline
& $\phi_+$ & $\tilde{H}_u$ & $H_d$ & ${\mathbf 10}$ & ${\mathbf 5}^*$& $\bar{N}_R $
& $\tilde{\Phi} $ & $\bar\Phi$ & Z
\\
\hline
$R$& $0$ & $4/5$ & $6/5$ & $3/5$ & $1/5$ & $1$& $4/5$& $8/15$& $2/3$
\\
\hline
$U(1)$& $+1$ & $0$ & $0$ & $0$ & $0$ & $0$& $+1$& $-1$& $0$
\\
\hline
\end{tabular}
\end{center}
\label{tab:charge}
\end{table}%

\section{Two-loop Contributions to Scalar Masses}\label{sec:appA}
Here, we show a few details of the calculation for the two-loop scalar masses. This calculation proceeds roughly the same as the case with $y_t'=0$.  The wave function renormalization is analytically continued into superspace so that it is a function of the spurion $X=M+\theta^2F$,

\begin{equation}
{\cal L}=\int d^4\theta {\cal Z}(|X|)\Psi^\dagger \Psi.
\end{equation}
The wave function renormalization can then be expanded in $\theta$ to give

\begin{eqnarray}
 {\cal Z}(|X|)= Z(M)+\left(\frac{\partial Z(M)}{\partial M}F\theta^2+\frac{\partial Z(M)}{\partial M^\dagger}F^\dagger\bar\theta^2\right)+\frac{\partial^2 Z(M)}{\partial M \partial M^\dagger}FF^\dagger \theta^4
 \end{eqnarray}
After the field rotation

\begin{equation}
\Psi\to Z^{-1/2}\left(1-Z^{-1}\frac{\partial Z(M)}{\partial M}\right)\Psi',\label{WavRenReg}
\end{equation}
we have the following Lagrangian
\begin{equation}
{\cal L}=\int d^4\theta ({\cal Z}(|M|)-\theta^4m_{\Psi}^2)\Psi^\dagger \Psi\ ,
\end{equation}
where
\begin{equation}
m_{\Psi}^2=\frac{1}{4}\left(\left(\frac{\partial Z}{\partial \ln M}\frac{1}{Z}\right)^2 -\frac{1}{Z}\frac{\partial^2 Z}{\partial^2 \ln M}\right)\frac{FF^\dagger}{MM^\dagger}\ ,\label{ScaMas}
\end{equation}
and we have used
\begin{equation}
\frac{\partial f(|M|)}{\partial \ln M}=\frac{1}{2}\frac{\partial f(|M|)}{\partial \ln |M|}\ .
\end{equation}

To this point, this calculation is the same as the standard calculation. For our model, however, this calculation is complicated by the one-loop kinetic mixing of the Higgs and messengers fields. These generated off-diagonal terms must be canceled by the wave-function renormalization. To simplify this calculation we initially break up our wave-function renormalization into two pieces: the one presented in Eq. (\ref{WavRenReg}) and the part that removes the kinetic mixing. To begin our discussion on the kinetic mixing, we show the generic form of the one-loop K\"ahler potential for the Higgs and messenger fields,

\begin{eqnarray}
K=\left(\begin{array}{cc}H_u^\dagger & \Phi^\dagger\end{array}\right)\left( \begin{array}{cc} 1+\delta Z_{H_u} & \delta Z_{H\Phi} \\
\delta Z_{H_u\Phi} & 1+\delta Z_{\Phi} \end{array}\right)\left(\begin{array}{c} H_u \\ \Phi\end{array}\right)
\ ,
\end{eqnarray}
where ${\cal Z}(|X|)=1+\delta Z_\Psi$. This matrix can be put in a diagonal, but not canonical, form by the field redefinition
\begin{eqnarray}
\left(\begin{array}{c} H_u'\\ \Phi'\end{array}\right) =\left( \begin{array}{cc} 1 & -\frac{1}{2}\delta Z_{H_u\Phi} \\
-\frac{1}{2}\delta Z_{H_u\Phi} & 1 \end{array}\right)\left(\begin{array}{c} H_u\\ \Phi\end{array}\right)\label{RotHPHi}
\end{eqnarray}
which gives
\begin{eqnarray}
K=\left(\begin{array}{cc}H_u^\dagger & \Phi^\dagger\end{array}\right)\left( \begin{array}{cc} 1+\delta Z_{H_u} & 0 \\
0 & 1+\delta Z_{\Phi} \end{array}\right)\left(\begin{array}{c} H_u \\ \Phi\end{array}\right)\label{WavRen}
\end{eqnarray}
to leading order. With this form of the Kahler potential, we can easily apply the techniques discussed above to calculate the two-loop contribution to the scalar masses. However, the rotation in Eq. (\ref{RotHPHi}) regenerates Higgs messenger mixing in the superpotential
\begin{equation}
\bar{g}Z\Phi\bar\Phi\to \bar{g}Z\Phi\bar\Phi-\bar{g}\frac{1}{2}\delta Z_{H_u\Phi} ZH_u\bar\Phi\ .\label{RegTer}
\end{equation}
If the field redefinition in  Eq. (\ref{RotHPHi}) is chosen to remove the kinetic mixing, applying the expression in Eq. (\ref{ScaMas}) is complicated. To return to a basis where these formulas can be simply applied, we make an additional unitary transformation of the Higgs and messenger superfields
\begin{eqnarray}
\left(\begin{array}{c} \hat H_u\\ \hat\Phi\end{array}\right) =\left( \begin{array}{cc} 1 &- \frac{1}{2}\delta Z_{H_u\Phi} \\
\frac{1}{2}\delta Z_{H_u\Phi} & 1 \end{array}\right)\left(\begin{array}{c} H_u'\\ \Phi'\end{array}\right)=\left( \begin{array}{cc} 1 & -\delta Z_{H_u\Phi}\\
0 & 1 \end{array}\right)\left(\begin{array}{c} H_u\\ \Phi\end{array}\right)\ .\label{WavMix}
\end{eqnarray}
If this new rotation is used to diagonalizes $Z$, we also find Eq. (\ref{WavRen}). The advantage of this rotation is it does not regenerate the problematic operator in Eq. (\ref{RegTer}). If we now combine the wave-function renormalization of Eq. (\ref{WavRenReg}) with that of Eq. (\ref{WavMix}), we have

\begin{eqnarray}
{\cal Z}_{tot}^{1/2}=\left( \begin{array}{cc} Z_{H_u}^{-1/2}\left(1-Z_{H_u}^{-1}\frac{\partial Z_{H_u}(M)}{\partial M}\right)& -\delta Z_{H_u\Phi}\\
0 & Z_\Phi^{-1/2}\left(1-Z_\Phi^{-1}\frac{\partial Z_\Phi(M)}{\partial M}\right) \end{array}\right)
\end{eqnarray}
for the wave-function renormalization of the Higgs and messenger fields. Since we will only considered contributions from the third generation, the wave-function renormalization for the other fields will be of the form in Eq. (\ref{WavRenReg}) and will not have any mixing.

Now we outline our procedure for calculating the derivatives of the wave-function renormalization.  First, the wave-function renormalization is formally solved for,

\begin{equation}
Z(M,\mu_R)=\int_{\ln M}^{\ln \mu_R}d\ln\mu~\gamma_L (\mu_R,M)Z(\mu_R,M)+\int_{\ln \Lambda}^{\ln M}d\ln\mu~\gamma_H (\mu_R)Z(\mu_R)\ .
\end{equation}
This expression is then differentiated. The derivatives are simplified using the expressions for the beta functions and we find

\begin{eqnarray}
m_\Psi^2=\frac{1}{4}\left(\frac{\partial \gamma_\Psi^L}{\partial \lambda^a}\Delta\beta_{\lambda^a}-\frac{\partial \Delta\gamma_\Psi}{\partial \lambda^a}\beta_{\lambda^a}^H\right)\frac{|F|^2}{M^2}\label{ScaMas2}
\end{eqnarray}
where
\begin{eqnarray}
\Delta\beta_{\lambda^a}=\beta_{\lambda^a}^H-\beta_{\lambda^a}^L\ ,\quad & \quad
\Delta\gamma_\Psi=\gamma_\Psi^H-\gamma_\Psi^L\ .
\end{eqnarray}

To find the sfermion masses in terms of our theories parameters, we need to determine the relevant beta functions and anomalous dimensions. The anomalous dimensions for the squarks, Higgs and messenger fields are

\begin{eqnarray}
&&\gamma_{H_u}^H=\frac{1}{32\pi^2}\left(6y_t^2-3g_2^2-\frac{3}{5}g_1^2\right),\cr
&&\gamma_{H_d}^H=\frac{1}{32\pi^2}\left(6y_b^2-3g_2^2-\frac{3}{5}g_1^2\right),\cr
&&\gamma_{\Phi}^H=\frac{1}{32\pi^2}\left(6y_t'^2-3g_2^2-\frac{3}{5}g_1^2\right),\cr
&&\gamma_{Q_3}^H=\frac{1}{32\pi^2}\left(2y_t'^2+2y_t^2+2y_b^2-\frac{16}{3}g_3^2-3g_2^2-\frac{1}{15}g_1^2\right),\label{gamma}\\
&&\nonumber\gamma_T^H=\frac{1}{32\pi^2}\left(4y_t'^2+4y_t^2-\frac{16}{3}g_3^2-\frac{16}{15}g_1^2\right),\\
&&\nonumber\gamma_B^H=\frac{1}{32\pi^2}\left(4y_b^2-\frac{16}{3}g_3^2-\frac{4}{15}g_1^2\right),\\
&&\nonumber\gamma_{H\Phi}^H=\frac{6y_ty_t'}{32\pi^2}\ .
\end{eqnarray}
The low scale anomalous dimensions can be found by taking $y_t'\to 0$.

The high-scale beta functions are found by applying
\begin{equation}
\beta_{\lambda}=\lambda\left(\gamma_a+\gamma_b+\gamma_b\right)
\end{equation}
where $\gamma_{(a,b,c)}$ are the anomalous dimensions of the fields for an interaction of the type $\lambda Q_aQ_bQ_c$. Using this expression, we find the following beta functions
\begin{eqnarray}
&&\nonumber\beta_{y_t}=\frac{y_t}{16\pi}\left(6y_t^2+3y_t'^2+y_b^2-\frac{16}{3}g_3^2-3g_2^2-\frac{13}{15}g_1^2\right),\\
&&\beta_{y_b}=\frac{y_b}{16\pi}\left(6y_b^2+y_t'^2+y_t^2-\frac{16}{3}g_3^2-3g_2^2-\frac{7}{15}g_1^2\right),\label{beta}\\
&&\nonumber\beta_{y_t'}=\frac{y_t'}{16\pi}\left(6y_t'^2+9y_t^2+y_b^2-\frac{16}{3}g_3^2-3g_2^2-\frac{13}{15}g_1^2\right),
\end{eqnarray}
where again the the low scale beta functions can be found by taking $y_t'\to 0$.  Combing the results of Eq. (\ref{ScaMas2}), Eq. (\ref{gamma}), and Eq. (\ref{beta}) we find the masses in Eq. (\ref{eq:Twoloop}),
\begin{eqnarray}
 \delta m_{Q_3}^2 &=& \frac{y_t'^2}{128\pi^4}\left(3y_t'^2 +3y_t^2
 - \frac{8}{3}g_3^2
 -\frac{3}{2} g_2^2
 -\frac{13}{30} g_1^2\right)\frac{F^2}{M^2}\ , \cr
 \delta m_{\bar T}^2 &=& \frac{y_t'^2}{128\pi^4}\left(6y_t'^2  + 6y_t^2
 + y_b^2
 - \frac{16}{3}g_3^2
 -3 g_2^2
 -\frac{13}{15} g_1^2\right)\frac{F^2}{M^2}\ ,\cr
  \delta m_{\bar B}^2 &=& - \frac{y_b^2y_t'^2 }{128\pi^4}\frac{F^2}{M^2}\ , \nonumber \\
  \delta m_{H_u}^2 &=& - 9\frac{y_t^2y_t'^2 }{256\pi^4}\frac{F^2}{M^2}\ , \nonumber \\
    \delta m_{H_d}^2 &=& - 3\frac{y_b^2y_t'^2 }{256\pi^4}\frac{F^2}{M^2}\ .\nonumber
\end{eqnarray}

\end{document}